\documentclass[preprint,12pt,longnamesfirst]{aastex}

\usepackage{latexsym}

\shorttitle{Massive YSOs in the LMC}
\shortauthors{Gruendl and Chu}

\newcommand {\ha}{H$\alpha$}

\begin{document}

\title{High and Intermediate-Mass Young Stellar Objects in the 
Large Magellanic Cloud}

\author{Robert A.\ Gruendl\altaffilmark{1,2} and You-Hua Chu\altaffilmark{1,2}}
\altaffiltext{1}{Astronomy Department, University of Illinois, 
        1002 W. Green Street, Urbana, IL 61801;
        gruendl@astro.illinois.edu,chu@astro.illinois.edu}
\altaffiltext{2}{Visiting astronomer, Cerro Tololo Inter-American Observatory}

\begin{abstract}
Archival {\it Spitzer} IRAC and MIPS observations of the Large Magellanic
Cloud (LMC) have been used to search for young stellar objects (YSOs).
We have carried out independent aperture photometry of these data and merged
the results from different passbands to produce a photometric catalog.  
To verify our methodology we have also analyzed the data from the SAGE and
SWIRE Legacy programs; our photometric measurements are in general agreement
with the photometry released by these programs.  A detailed completeness 
analysis for our photometric catalog of the LMC show that the 90\% completeness 
limits are, on average, 16.0, 15.0, 14.3, 13.1, and 9.2 mag at 3.6, 4.5, 
5.8, 8.0, and 24~$\mu$m, respectively.

Using our mid-infrared photometric catalogs and two simple selection 
criteria, [4.5]$-$[8.0]$>$2.0 to exclude normal and evolved stars and 
[8.0]$>$14$-$([4.5]$-$[8.0]) to exclude background galaxies, 
we have identified a sample of 2,910 sources in the LMC that could 
potentially be YSOs.  We then used the {\it Spitzer} observations 
complemented by optical and near-infrared data to carefully assess 
the nature of each source.  To do so we simultaneously considered 
multi-wavelength images and photometry to assess the source morphology, 
spectral energy distribution (SED) from the optical through the 
mid-infrared wavelengths, and the surrounding interstellar environment 
to determine the most likely nature of each source.

From this examination of the initial sample, we suggest 1,172 sources 
are most likely YSOs.  We have also identified 1,075 probable background 
galaxies, consistent with the expected number estimated from the SWIRE survey.
{\it Spitzer} IRS observations of 269 of the brightest YSOs from our
sample have confirmed that $\gtrsim$95\% are indeed YSOs.
Examination of color-color and color-magnitude diagrams shows no simple 
criteria in color-magnitude space 
that can unambiguously separate the LMC YSOs from all AGB/post-AGB stars, 
planetary nebulae, and background galaxies.

A comprehensive search for YSOs in the LMC has also been carried out by
the SAGE team and reported by \citet{Wetal08}.  There are three major 
differences between these two searches.  (1) In the common region of 
color-magnitude space, $\sim$850 of our 1,172 probable YSOs are missed
in the SAGE YSO catalog because their conservative point source 
identification criteria have excluded YSOs superposed on complex stellar
and interstellar environments. (2) About 20--30\% of the YSOs identified
by the SAGE team are sources we classify as background galaxies.
(3) the SAGE YSO catalog identifies YSO in parts of color-magnitude space 
that we excluded and thus contains more evolved or fainter YSOs missed by our
analysis.
The shortcomings and strengths of both these YSO catalogs should 
be considered prior to statistical studies of star formation in the LMC.
Finally, the mid-infrared luminosity functions in the IRAC bands of our 
most likely YSO candidates in the LMC can be well 
described by $N(L) \propto L^{-1}$, which is consistent with the Salpeter 
initial mass function if a mass-luminosity relation of $L \propto M^{2.4}$ 
is adopted.
\end{abstract}

\keywords{Magellanic Clouds --- stars: formation --- surveys --- infrared: general}

\section{Introduction}

Star formation is one of the most fundamental processes that shape the
observable universe.  In particular, the formation of massive stars
dramatically alters their local environment as their strong UV
radiation field, stellar winds and eventual explosions as supernovae
inject energy into the surrounding interstellar medium (ISM).
This stellar energy feedback may compress the ambient medium to
induce subsequent star formation, but may also disperse the natal
molecular clouds to regulate further star formation.  The onset
and propagation of star formation plays an important role in
the evolution of a galaxy.

Investigations of the star formation process using individually
resolved young stars have been conducted for regions in our Milky
Way galaxy.  The formation of low-mass stars can be studied in
great detail in nearby star forming regions \citep[e.g., Taurus-Auriga
Molecular Cloud;][]{KH95,Hetal05}, but the formation of stars more massive than
a few solar masses cannot be studied easily because such stars are
rarer and are often found in distant regions toward which the line-of-sight
obscuration and confusion in the Galactic plane limits our ability
to obtain a clear view.  A global view of star formation in the
Galaxy is particularly impossible.

The {\it Spitzer Space Telescope}, with its high angular resolution
and sensitivity at mid- and far-infrared (IR) wavelengths, has provided
not only a better view of the formation of individual massive and
low-mass stars in the Galaxy, but also a new opportunity to study
massive star formation in nearby galaxies, most notably the Large
Magellanic Cloud (LMC). In the LMC, because of its close proximity
\citep[50~kpc, where 1\arcsec =0.25~pc;][]{F99} and low inclination
\citep[$\sim30^\circ$;][]{Netal04}, massive and intermediate-mass
young stellar objects (YSOs) can be resolved by {\it Spitzer} and
inventoried throughout the entire galaxy.  

In this paper we consider YSOs to be young stars still in 
the process of forming, as YSO candidates selected based on
mid-IR {\it Spitzer} observations have already formed a compact 
source at the center.  Furthermore, the linear resolution of 
{\it Spitzer} observations, $>$50,000~AU, means that a central source and its
surrounding circumstellar material cannot be separated.  Thus,
the emission would include not only the central source but also
any circumstellar disk, and/or circumstellar envelope.  In terms of
Class 0/I/II/III system used to describe low-mass YSOs \citep{Lada87}, 
we expect to be biased toward high and intermediate-mass systems that 
are more similar to the Class I and II sources.  In other words, sources
that have formed a central source but which may still be in the process
of actively accreting material.  As such, the central objects are most 
likely young pre-main sequence stars.

This inventory of massive and intermediate-mass YSOs,
combined with the well-surveyed ISM and stars, can be used to study
the relationship between star formation and gravitational instability
on a global scale \citep[e.g.,][]{Yetal07}, as well as triggered
star formation on local scales \citep[e.g.,][]{Cetal05,Cetal08,Chen09}.
It is also possible to investigate whether the mass function
of the newly formed stars depends on the interstellar conditions that
lead to the star formation \citep[e.g.,][]{CG08}.

In this paper we present the results of a search for YSOs in the LMC
using {\it Spitzer} observations and complementary optical and near-IR
observations.  In \S{2} we describe the observations used for this
search.  In \S{3} we examine the mid-IR photometry, compare our
results and methodology with those from two other wide-area surveys
using {\it Spitzer}, and examine the completeness and reliability of
our results.  In \S{4} we describe the methodology to search
for YSOs using the photometric results and in \S{5} we present the
results of this search.  Finally, in \S{6} we discuss the populations
of sources that we identify, compare them with the results from a 
previous YSO survey, and discuss the implications of these 
results for studying the detailed star formation in the LMC.

\section{Observations and Data Reduction}

\subsection{Data Set}

The LMC has been observed
by the {\it Spitzer Space Telescope} using the InfraRed Array Camera 
\citep[IRAC;][]{Fetal04} and the Multiband Imaging Photometer for {\it Spitzer} 
\citep[MIPS;][]{Retal04}.  We have used the archival data from the 
{\it Spitzer} Legacy program Surveying the Agents of a Galaxy's 
Evolution \citep[SAGE;][]{Metal06} that mapped the central 7$^\circ\times$7$^\circ$ 
area of the LMC, along with numerous earlier programs that targeted 
star-formation complexes throughout the LMC (see Table~\ref{observe_tab} for 
summary).
We have downloaded both the BCD (Basic Calibrated Data) and post-BCD pipeline
processed products for each of the programs listed as they became available.
More information on the instruments and pipeline processing can be found at 
the {\it Spitzer} Science Center's Observer Support 
website\footnote{http://ssc.spitzer.caltech.edu/ost.}. 

\subsection{IRAC Data Reduction}

For the IRAC observations we have used the post-BCD products to make 
photometric measurements of all sources in the IRAC 3.6, 4.5, 5.8, 
and 8.0~$\mu$m bands.  The DAOFIND task in IRAF was used to 
search for sources in the IRAC post-BCD images with the {\it sharpness} 
and {\it roundness} criteria optimized to: account for the point-spread 
function (PSF) afforded by IRAC, aid in the 
rejection of cosmic-ray hits, and identify sources amid a complex 
background.  We found the ranges for these two criteria to identify
the largest numbers of reliable sources were: $0.6 \ge sharpness \ge 1.0$ 
and $-0.7 \ge roundness \ge 0.7$.
The PHOT task was then used to obtain photometric extractions from 
a 3\farcs6 radius aperture centered on each source and a 3\farcs6--8\farcs4 
annular background region.  To determine
source brightnesses we then applied the aperture corrections and flux calibrations 
tabulated in the IRAC Data Analysis Handbook ver2.0.  When observations 
were obtained with the high dynamic range
mode, photometric measurements were also extracted from the short-exposure 
images to obtain measurements for sources at or near saturation in the long 
exposures.  In Table~\ref{phot_tab} we summarize the photometric parameters, 
aperture corrections, zero-points, and the assumed flux 
calibration accuracy used when analyzing the IRAC observations.

At each location in the LMC there are usually IRAC 
observations for at least two epochs.  The photometric results in each band
were compared and combined where individual measurements were weighted by 
the inverse square of their photometric uncertainty.  
The post-BCD images contain numerous transient sources,
including cosmic-ray hits, ghost images of bright sources, and latent images 
of bright sources from prior frames.  To remove the transient sources
we first relied on the above-mentioned {\it sharpness} and {\it roundness} criteria 
to reject sources that are significantly more peaked or less circular 
than the IRAC PSF.  Further rejection of transients was accomplished when combining 
data from different epochs by eliminating ``sources" that should have had 
statistically significant counterparts at other epochs.  An assessment of
the resulting photometric reliability and completeness for the IRAC measurements
are made in \S~\ref{phot_check}.  

In order to match sources between different IRAC bands we then 
cross-correlated the locations of sources between bands.  This was accomplished
by an algorithm which first intercompared each pair of IRAC bands singly and then
reconciled those sets of comparison with one another.  For the IRAC
measurements, positions were considered to match if they were within 1\farcs5. 
This process often results in ``orphan'' sources that have a detection in only
one band (most typically the 3.6~$\mu$m band).  On rare occasions a ``degenerate''
match can occur, where a source in one band is matched to multiple sources 
in another band.  For these cases the closest match was assumed to be correct.
The resulting photometric catalog has $>$3.5 million IRAC sources where the
position of each source is formed by the weighted mean of the positions found
in the individual bands.

\subsection{MIPS Data Reduction}

Most of the MIPS observations in the LMC have used the scan map mode
with either a medium or fast scan rate.  For the MIPS 24 and 70~$\mu$m 
observations we used the MOPEX software to construct new mosaics 
from all available BCD data with 1\farcs245 and 4\farcs0~pixels 
($\sim\frac{1}{2}$ the original resolution), respectively.
Prior to building each mosaic we removed brightness offsets
between the individual frames by solving for the best offset for each 
frame using the method outlined in \citet{RG95}.  In this image reconstruction 
we relied on the fact that the SAGE MIPS observations usually had at least 
two scans at nearly orthogonal directions to remove most of the bright 
latent images that occurred after the MIPS detectors encounter bright 
emission.  Some artifacts could remain in these images.

In the case of the MIPS 24~$\mu$m reconstructed images we again searched for 
sources using the DAOFIND task and then performed aperture photometry 
using the PHOT package in IRAF.  The photometric extractions 
used a 6\arcsec\ radius aperture centered on the source with a 15-23\arcsec\ 
annular region to measure the background flux.  These
aperture sizes were chosen to allow us to analyze sources with as small a
separation as possible while also minimizing background contamination by
placing the sky aperture between the first and second Airy rings.  We
used isolated bright sources to independently determine the aperture correction 
and found a value of 1.798.  Our value is high compared to the typical 
values of 1.69 tabulated in the MIPS Data Analysis Handbook (versions 3.2 and 
earlier), but recent work by \citet{Fetal06} measuring the
curve of growth for MIPS 24~$\mu$m observations results in aperture corrections 
of 1.78 and 1.84 for a theoretical PSF and an empirical determination from 
the {\it Spitzer} Extragalactic First Look Observations, respectively.
We have used our aperture correction to determine source brightnesses at 24~$\mu$m 
throughout this work.  

Similar to the IRAC observations, there exist MIPS 24~$\mu$m observations
at two or more epochs for nearly all locations.  Thus transients not 
rejected by MOPEX are mostly rejected when we combine all 24~$\mu$m 
photometric measurements using the same methods and criteria used for 
the IRAC measurements.  The final catalog at 24~$\mu$m has 53,800 sources.
These source locations have been cross-correlated with and incorporated 
into the merged IRAC catalog for 24~$\mu$m sources within 1\farcs5 of an 
IRAC source.  Here multiple matches are not possible because the MIPS~24~$\mu$m 
PSF has a full-width at half maximum of $\sim$6\arcsec.
An assessment of the resulting photometric reliability and completeness for the 
MIPS~24~$\mu$m measurements are made in \S~\ref{phot_check}.  

In the case of the MIPS 70~$\mu$m observations, rather than a blind search 
for sources using the DAOFIND task, we have made photometric 
measurements for two situations.  First, we measure fluxes for 70~$\mu$m 
sources where we have identified an unambiguous counterpart in the 70~$\mu$m 
images for sources previously identified with IRAC and/or MIPS~24~$\mu$m.
Second, where no 70~$\mu$m source is present but an upper-limit 
may help constrain the nature of a source identified at other wavelengths,
we measure the background variation and determine a 3-$\sigma$ upper-limit.
Both types of photometric extractions were made using the PHOT package 
in IRAF where we used a 18\arcsec\ radius source aperture and an annular
background region of radii 18\arcsec -- 39\arcsec.  Based on the results in the 
MIPS Data Analysis Handbook 
(version 3.2) we used an aperture correction of 1.927 to obtain our final 
source brightnesses.  The photometric extraction parameters for the MIPS 
observations are also given in Table~\ref{phot_tab}.

\subsection{Complementary Optical and Near-IR Observations}\label{sec_obsoptnir}

We have obtained complementary optical and near-IR imaging observations 
using the Blanco 4m telescope at the Cerro Tololo Inter-American Observatory 
for selected regions throughout the LMC.  The MOSAIC\,{\sc ii} camera was used 
to obtain $I$-band observations on 2006 February 2--8. 
The MOSAIC\,{\sc ii} camera has a 36\arcmin$\times$36\arcmin\ field-of-view
that is imaged by eight SITe 4096$\times$2048 CCDs with 0\farcs27 pixels.  
A more detailed description of this camera can be found in \citet{Metal98}.
Seven fields were observed, with five centered on the \ion{H}{2} regions 
N44, N51, N70, N144, and N180 \citep{Henize56} and the remaining two covering
the northern half of the supergiant shell LMC\,3 \citep{GM78,Meaburn80}.
Each field was imaged with three 120~s exposures, along with single 1~s 
and 10~s exposures.  Small pointing offsets between exposures were made so
that the combined images would cover the gaps between CCDs.  The data reduction
used the SuperMACHO pipeline software and included bias subtraction, 
flat fielding, and distortion correction 
\citep[for more details see][]{Setal02}.  These observations have an angular 
resolution $\lesssim$~1\farcs0 and detect point sources of 
m$_I \lesssim$22.5 mag with a 10-$\sigma$ significance or better.

The IR Side Port Imager \citep[ISPI;][]{vdBetal04} was used to obtain 
$J$ and $K_s$-band observations of 85 fields
throughout the LMC during three observing runs in 2005 November, 2006 November,
and 2007 February.  The ISPI camera has a 10\farcm25$\times$10\farcm25
field-of-view imaged with a 2048$\times$2048 HgCdTe Array with 0\farcs3 pixels.
Each field was imaged with a sequence of exposures with small ($\sim$1\arcmin) 
telescope offset between frames to aid in removal of bad pixels and to 
facilitate sky subtraction and flat fielding.  For the $J$-band observations
a sequence of thirteen 30~s exposures were obtained while at $K_s$-band a 
sequence of twenty-three 30~s exposures (each consisting of two coadded
15~s exposures) were obtained.  The observations were non-linearity corrected,
sky subtracted, and flat fielded using standard routines within the CIRRED
and SQIID packages in IRAF.  An astrometric frame for each exposure was then
established using the WCSTOOLS program IMWCS and the Two Micron
All Sky Survey Point Source Catalog \citep[2MASS PSC;][]{Setal06}.  
Prior to mosaicing the exposures together the relative brightness offset 
in the background was solved for and removed from each exposure using the 
methods outlined by \citet{RG95}.  When mosaicing the exposures typically 
the first and last exposures were dropped due to poor background subtraction.  
The resultant
mosaiced images have a typical effective exposure time of $\sim$300~s
and 600~s at $J$ and $K_s$-bands, respectively.  The images were flux 
calibrated by bootstrapping from stars in common with the 2MASS PSC
and the typical accuracy achieved ranged between 3 and 7\%.  The angular 
resolution of the final mosaics is typically $\lesssim$~1\farcs0 and 
point sources with m$_J\lesssim$18.5 and m$_{K_s}\lesssim$17.6 mag are
generally detected with better than 10-$\sigma$ significance.

\section{Mid-IR Photometric Consistency Checks: Accuracy and Completeness}\label{phot_check}

To validate our method of photometric extraction we have downloaded and
analyzed both the IRAC and MIPS 24~$\mu$m data from the {\it Spitzer} Wide-area 
Infrared Extragalactic Survey \citep[SWIRE;][]{Letal03} and compared our 
photometric measurements with those of the SWIRE team (data releases DR2 and DR3).
When the SAGE photometry of the first epoch data was released, we
further compared our photometric measurements with those from the
catalogs SAGEcatalogIRACepoch1 and SAGEcatalogMIPS24epoch1 (hereafter SAGE DR1).

\subsection{Photometric Accuracy}\label{sec_accuracy}

\begin{figure}
\epsscale{0.9}
\plotone{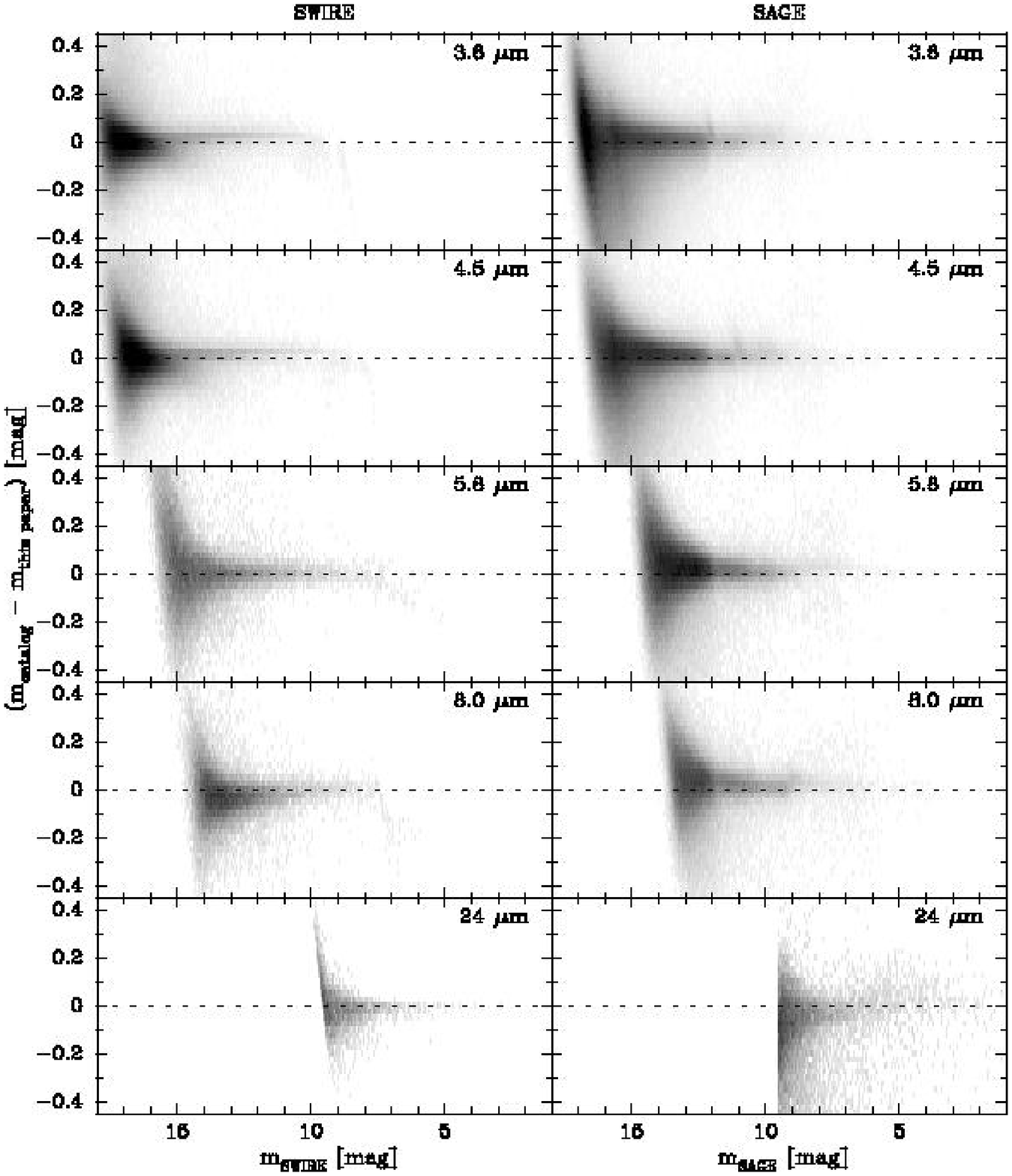}
\caption{Comparison with photometric results from the SWIRE and SAGE catalogs.
The horizontal axis of the panels in the left column use the measurements 
from the SWIRE DR2 and DR3 catalogs while those in the right column use the
measurements from the SAGE DR1 catalog.  The vertical axis shows the 
difference between the catalog values and our measurements.}
\label{photoacc_fig}
\end{figure}

The results from comparison of our photometric measurements with those of 
the SWIRE and SAGE surveys are shown in Figure~\ref{photoacc_fig}.  We find 
reasonable agreement with both surveys.  The SWIRE survey has used the 
SExtractor software \citep{BA96} to obtain their photometric measurements.  
None of the apertures used for source extraction by the SWIRE team are 
identical to those used for our LMC work so we have chosen to make a 
comparison with their results obtained with a slightly larger aperture (4\farcs1
vs. our 3\farcs6 aperture) to minimize any effect that might arise from slight 
differences in the aperture locations relative to the source centroid.

The photometric measurements released by SAGE were obtained by using 
a modified version of the IRAF routine DAOPHOT \citep{S87} that 
was developed to process data of the GLIMPSE Legacy Program \citep{Betal03}.
Similar to this work sources were initially identified using DAOFIND, however,
less restrictive sharpness and roundness criteria were used 
($0.2 \ge sharpness \ge 1.2$ and $-1.5 \ge roundness \ge 1.5$; private 
communication B. Babler).  Cosmic-ray hits were rejected in subsequent 
processing steps (a more detailed description is available on
the GLIMPSE and SAGE websites: http://www.astro.wisc.edu/glimpse and 
http://sage.stsci.edu/).  Unlike the simple aperture extraction used for both 
SWIRE and this work, the SAGE analysis uses PSF-fitting which enables photometric 
estimates even in crowded fields.  

In order to further quantify these comparisons for each of the IRAC bands
and the MIPS 24~$\mu$m band, Table~\ref{compphot_tab} lists:
N$_\lambda$, the number of measurements in common with SWIRE and SAGE;
$\Delta_{\lambda}$, the median offset of our measurements from those made 
by SWIRE and SAGE in magnitudes; and $\sigma_{\lambda}$, the root mean 
squared (rms) dispersion around the median offset in magnitudes. 
Our results show small systematic offsets of a few hundreths of a
magnitude when compared to those obtained by SWIRE and SAGE.  These
systematic offsets appear to grow larger as we approach and exceed
our completeness limits (see \S~\ref{sec_complete}).  Our results appear
to better match those of SWIRE, as the comparison with the SAGE
results generally have a larger dispersion. This is not surprising as 
the SWIRE photometric extraction method is similar to ours and the SWIRE
observations generally do not suffer from crowding.  Based on the results in
Figure~\ref{photoacc_fig} and Table~\ref{compphot_tab} differences 
as large as 20\% between our flux measurements and those made by 
the SAGE team could be expected; however, our photometric results should 
be adequate to identify YSOs.

\subsection{Completeness and Crowding}\label{sec_complete}

\begin{figure}
\plotone{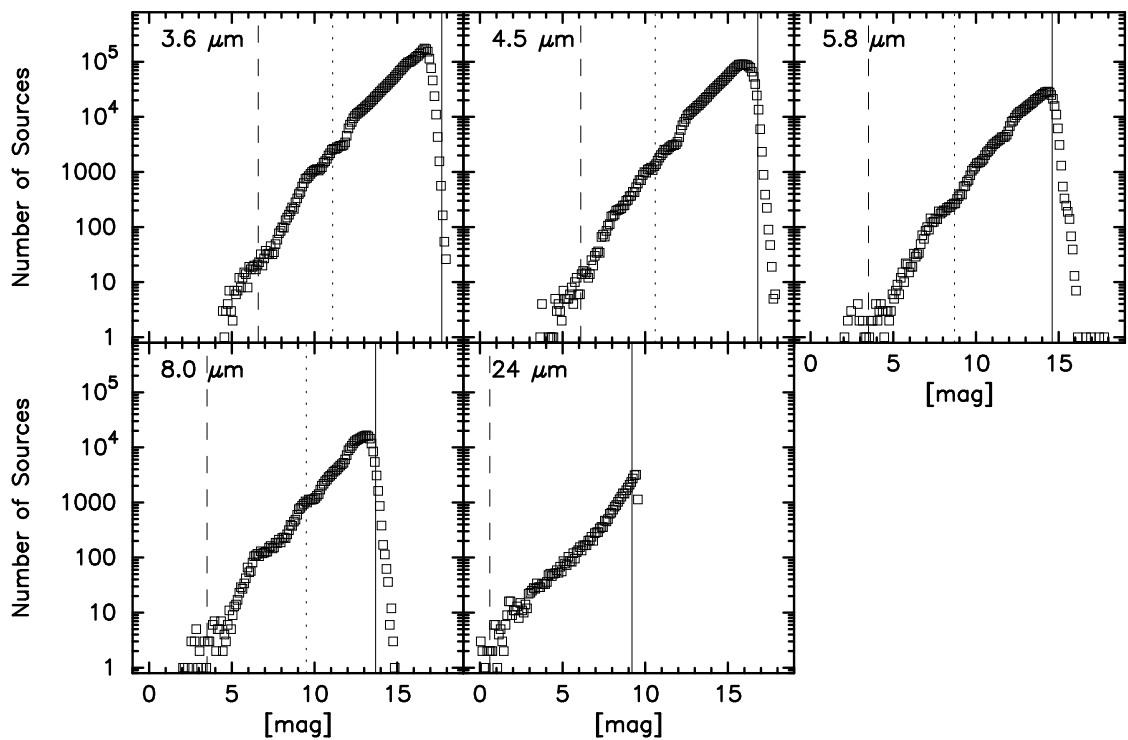}
\caption{Luminosity functions constructed using our IRAC and MIPS 24~$\mu$m 
measurements for the entire LMC survey area.  In each panel a dashed line
marks the brightness where saturation should begin to significantly affect the 
photometry of bright sources.  A dotted line is used to indicate the rough brightness 
where we switch between the long and short exposures taken in the high dynamic range
mode.  A solid line marks the the expected brightness for a point source 
to be detected with 10$\sigma$ significance using the SAGE survey data.}
\label{lumfunc_fig}
\end{figure}

We have constructed luminosity functions for each IRAC and MIPS band for the 
entire LMC (see Figure~\ref{lumfunc_fig}). We find that our source counts peak 
at 16.6, 16.0, 14.3, 13.1, and 9.4 mag for 3.6, 4.5, 5.8, 8.0, and 24~$\mu$m, 
respectively.  To estimate the completeness limits for our IRAC and MIPS 24~$\mu$m 
photometry we have randomly placed test sources with varying brightness throughout the 
original IRAC and MIPS post-BCD images.  We then searched for these test sources
using the same software that was used for our original search.  The test sources
were created using the IRAC and MIPS point response functions (PRFs) with 1/5 pixel 
sampling available on the {\it Spitzer} Science Center website.  
For the IRAC, the PRF shape changes as a 
function of its position on the array and 25 different PRFs are available.
In order to accommodate this varying PRF we alternated among the 25 IRAC PRFs 
using a different PRF for each test source.  When adding a test source to an 
image we also include Poisson noise to reflect the photon statistics for the new 
source.  We note that this additional noise is a slight over-estimate but should 
only make a minor difference except for the faintest test sources.  

The SAGE survey mapped the central 7$^\circ\times$7$^\circ$ area of the LMC
at two epochs.  At each epoch the IRAC portion of the survey is composed of 49 
fields that are $\sim$1.35 square degrees each.  Similarly, the MIPS~24~$\mu$m 
portion of the survey at each epoch is composed of 38 strips, taken in the scan 
map mode, which have a width of $\sim$27\arcmin\ and cover $\sim$2.05 square degrees 
each.  These observations produced 98 post-BCD images for each IRAC band and 78 
post-BCD images for each MIPS band and make up the majority of the observations 
used in our analysis.

We randomly placed 1000 test sources with a constant brightness in each of the
post-BCD images from the SAGE survey.  Only 1000 sources were added
at a time so that the number density of sources in the image would not change
significantly.  We then used the same IRAF scripts to identify and extract 
sources.  The results were then checked to determine whether or not the test 
sources were detected.  Our software which merged the individual photometric 
catalogs from each post-BCD image made cross-checks among the resulting lists.  
These checks were designed to both remove cosmic-ray hits and to double check 
sources that were not identified in all possible observations.  To test this 
portion of our algorithm, test sources in an overlapping region of multiple images
are placed at the same sky positions in each of these images.  A test source 
was deemed ``found" or ``recovered" if it was found in any of the images and not 
rejected by the cross-check.  This tends to increase our detection rate near 
the completeness limit because marginal cases have multiple chances to be identified
and then reconfirmed in the cross-check stage.

The entire process was repeated for sources with brightnesses spanning the
range of those detected in the survey and was repeated for all the IRAC 
and MIP 24~$\mu$m images that made up the SAGE survey.  This results in a 
total of 98,000 test sources for each brightness level in each of the IRAC bands
and 78,000 test sources in the MIPS 24~$\mu$m band.  These trials made up the 
main body of our search and enables an estimate of the overall completeness 
trends when considering the entire LMC.  

In order to investigate the effects of crowding and bright diffuse emission 
on the completeness limits, we recorded the number of sources within 1\arcmin\ 
of each test source and the RMS of the background prior to adding the test 
source.  Note that a direct comparison to the local background is not possible 
because the photometry was carried out using post-BCD products which contain 
background offsets and gradients.  To have sufficient statistics to make a 
meaningful comparison in regions with either a high number density of sources 
and/or bright diffuse emission it was necessary to make additional tests using 
a few images that contained these rarer conditions.

\begin{figure}
\plotone{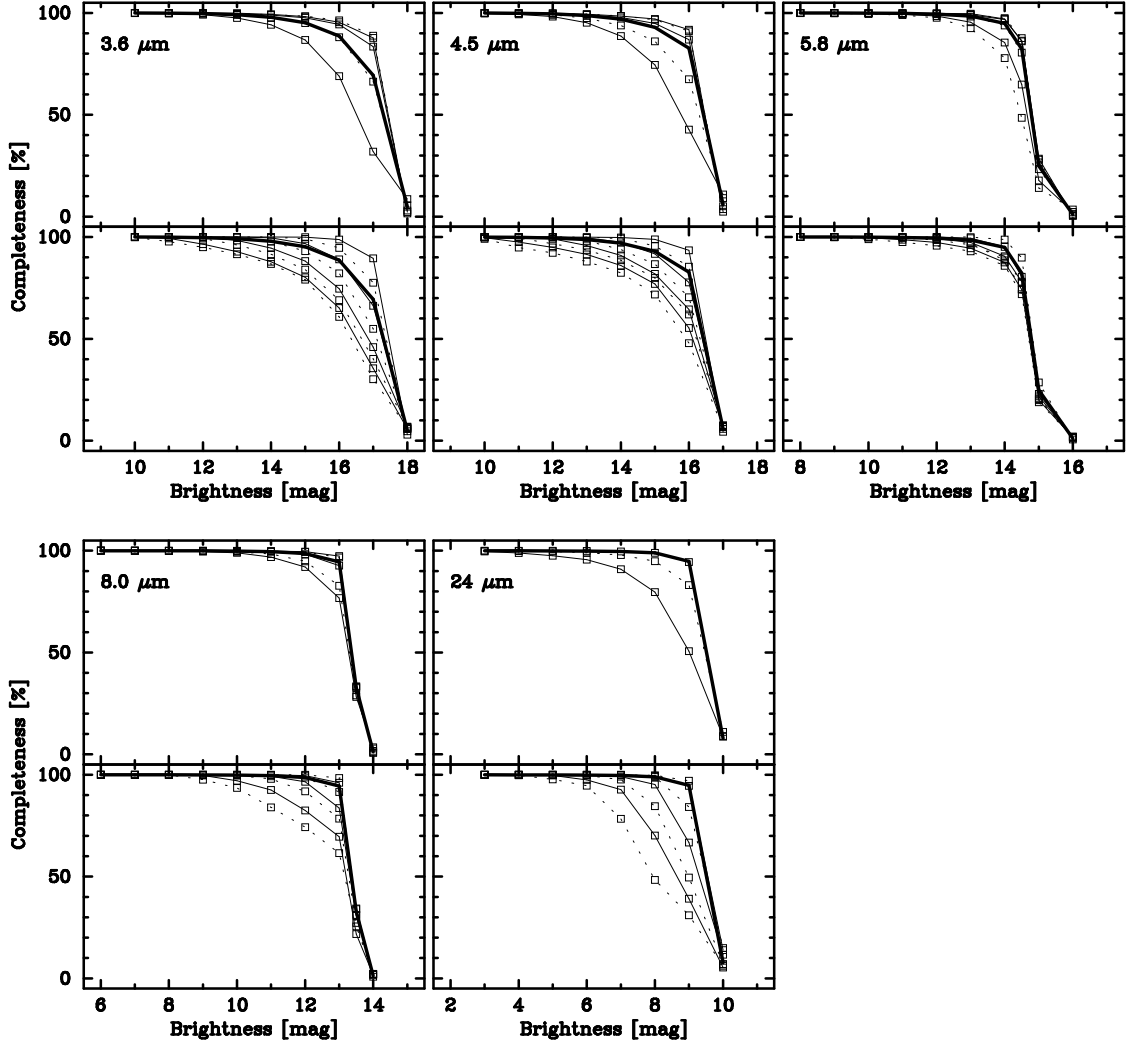}
\caption{The results from our detailed completeness tests.  For each band
two panels are shown plotting completeness vs. source magnitude.  The heavy 
solid line in each panel indicates the average completeness when considering 
the entire LMC survey area equally.
The top panels for each band show completeness as a function of crowding
while the bottom panels show completeness as a function of RMS background.
The alternating solid and dotted lines each represent a factor of 2 increase
in number density of sources (ranging from 4--32~arcmin$^{-2}$ at 3.6 and 
4.5~$\mu$m, 1--16~arcmin$^{-2}$ at 5.8~$\mu$m, 1--8~arcmin$^{-2}$ at 8.0~$\mu$m,
and 1--4~arcmin$^{-2}$ at 24~$\mu$m) or a factor of 2 increase in RMS 
background (ranging from 0.05--6.4~MJy~sr$^{-1}$ at 3.6 and 4.5~$\mu$m,
0.1--6.4~MJy~sr$^{-1}$ at 5.8, 8.0 and 24~$\mu$m).}
\label{comp_stat}
\end{figure}

Figure~\ref{comp_stat} shows the results of our analysis of the completeness limits.
When considering the LMC survey region as a whole the 90\% completeness limits
are 16.0, 15.0, 14.3, 13.1, and 9.2 mag at 3.6, 4.5, 5.8, 8.0, and 24~$\mu$m,
respectively.  These values are similar to or slightly brighter than those
implied by the peaks in the source luminosity functions (Figure~\ref{lumfunc_fig}).  
In the most crowded regions, those with a source density of $>$24~arcmin$^{-2}$ at 
3.6 and 4.5~$\mu$m, the completeness limits drop by roughly 1~mag.  Less than 
0.5\% of the survey region has such number densities.  At 5.8, 8.0, and 24~$\mu$m the
source number densities rarely exceed 16, 8, and 4~arcmin$^{-2}$, respectively.
Figure~\ref{comp_stat} appears to suggest that even such low number densities 
change the completeness limits at 24~$\mu$m, but this really results from 
higher background emission which generally occurs in star forming regions where
the 24~$\mu$m source number densities are highest.

\begin{figure}
\plotone{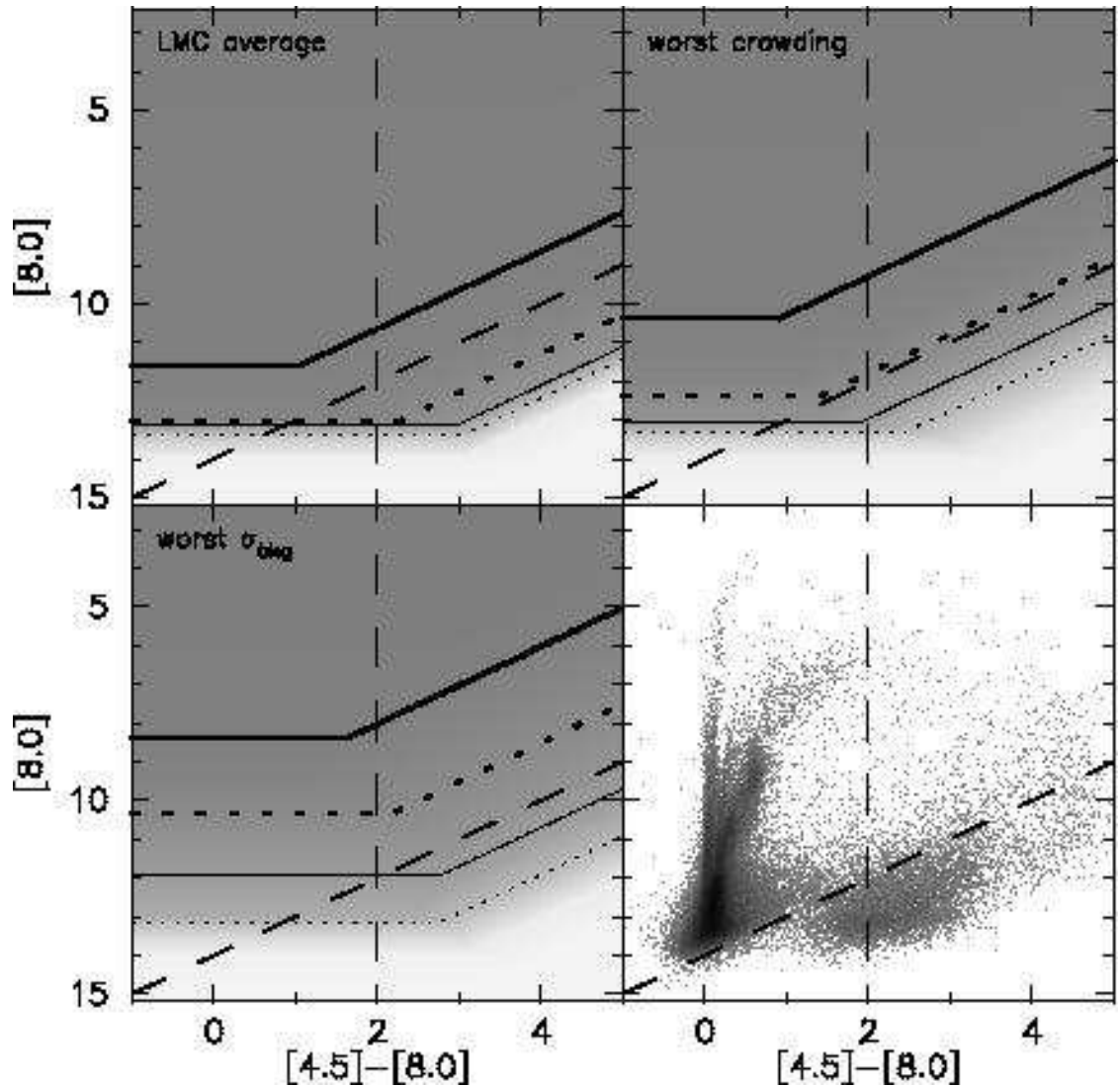}
\caption{Completeness results in the initial color-magnitude space used
to select YSO candidates.  The thick solid lines mark the limit for 99\%
completeness while the thick dotted, thin solid, and thin dotted lines
mark the limits for 90\%, 75\%, and 50\% completeness, respectively.  
The heavy dashed lines mark the cutoffs that define our initial YSO 
sample.  The {\it top left} panel shows the completeness limits when 
considering the entire LMC survey area.  The {\it top right} and {\it bottom 
left} panels show the completeness limits under the ``worst" crowding and 
background conditions we were able to probe.  The {\it bottom right} panel 
shows the [8.0] vs. [4.5]$-$[8.0] Hess diagram of the LMC for comparison.}
\label{comp_YSO}
\end{figure}

Examination of the completeness results as a function of the RMS background
shows the extent to which a higher background will impair our ability to identify
sources.  At the highest RMS background levels that we could probe statistically
($\sim$5--10~MJy~sr$^{-1}$) we found that the 90\% completeness limits dropped by 
$\sim$3~mag.  While these regions again make up less than 0.5\% of the survey 
region they are most likely to occur amid bright dust emission and therefore may
be associated with the YSO candidates.  Figure~\ref{comp_YSO} demonstrates how 
the completeness limits compare with the color-magnitude space used to identify
our initial sample of YSOs (see \S~\ref{ysocand}).  We estimate that
this initial YSO sample will begin to become incomplete for [8.0]$>$8.0.

\begin{figure}
\epsscale{1.0}
\plotone{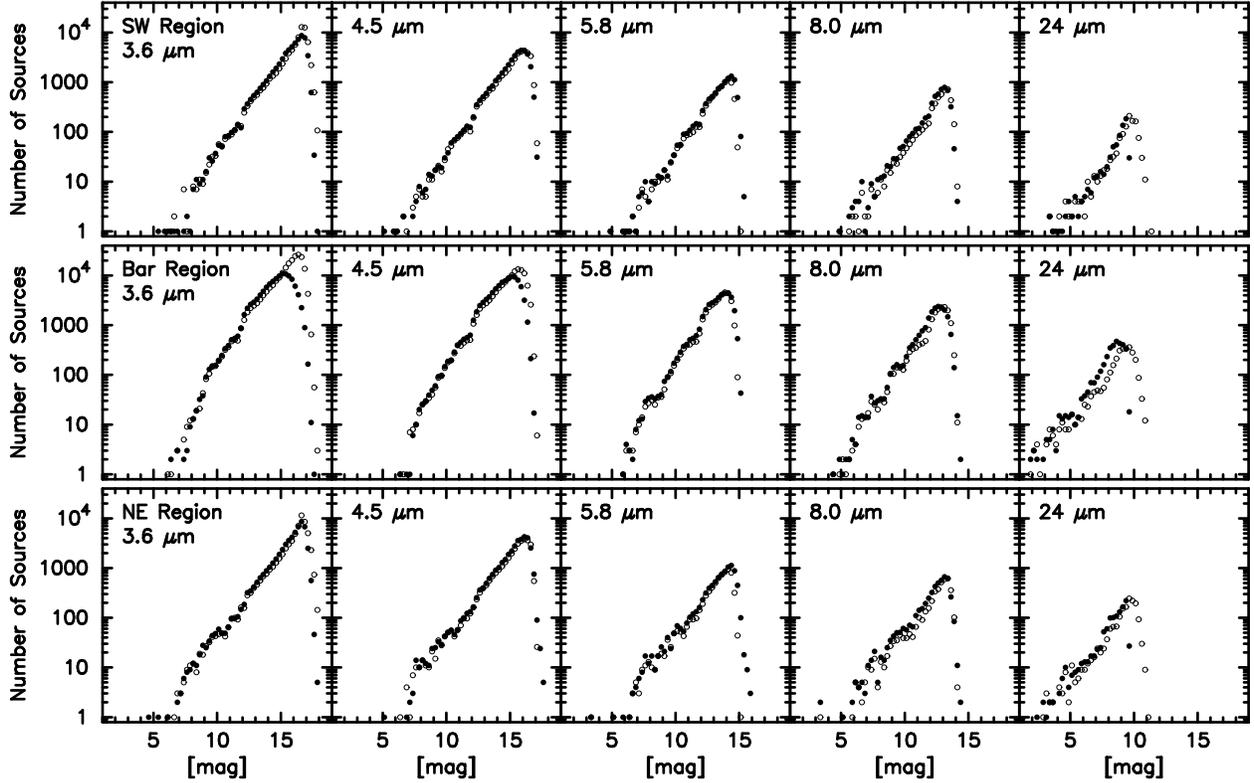}
\caption{Luminosity functions for the IRAC and MIPS 24~$\mu$m bands from 
three $\sim$1~square degree regions in the LMC.  Results from our photometry 
are shown as filled circles while those from the SAGE DR1 release are shown 
with open circles.  The SW region ({\it top panels}), Bar region ({\it center 
panels}), and NE region ({\it bottom panels}) are centered at 
4$^h$58$^m$,$-$70\arcdeg 30\arcmin\, 5$^h$21$^m$,$-$69\arcdeg 30\arcmin\, and 
5$^h$34$^m$,$-$66\arcdeg 30\arcmin\, respectively.  The SW and NE regions 
have a lower density of sources and less mid-IR nebular emission 
when compared to the Bar region.}
\label{crowding_fig}
\epsscale{1.0}
\end{figure}

To further investigate whether our aperture photometry may suffer
some incompleteness for sources closer than $\sim$3\arcsec\ due to crowding;
we have compared our luminosity functions for three $\sim$1\arcdeg$\times$1\arcdeg\
regions in the LMC with those constructed from the SAGE DR1, which should
suffer less from crowding since the photometry was obtained using PSF fitting.
The three regions are located (1) in the north-east portion of the LMC covering the northern
half of the supergiant shell LMC-4 which contains numerous stellar associations,
(2) in the LMC bar at a position where the highest surface density of
sources is seen, and (3) in the south-west portion of the LMC in a modest
stellar density environment. 
The comparisons, in Figure~\ref{crowding_fig}, show that in most regions our 
source counts are roughly similar to those found in the SAGE DR1, but in 
regions such as the LMC bar where crowding becomes significant we may miss 
many faint sources, particularly at 3.6 and 4.5~$\mu$m.  Furthermore, in the LMC 
bar region we find a clear discrepancy at 24~$\mu$m where the luminosity 
functions generated from our photometry and the SAGE DR1 appear shifted 
in brightness by roughly 1 magnitude with respect to each other.  

In order to better understand the discrepancy between our 24~$\mu$m
photometry and those in the SAGE DR1 results, we have examined the 
location of sources as a function of the difference between our photometry 
and the SAGE results.  We find that the sources with larger differences
are generally located in regions with significant diffuse 24~$\mu$m emission.
Furthermore, the larger differences occur for sources projected amid the 
brightest diffuse emission or within filamentary 24~$\mu$m emission.  In such
cases our background annulus (between 15-23\arcsec\ radius, to avoid
the Airy rings in the MIPS 24~$\mu$m PSF) could often under estimate the 
background level resulting in an overestimate for the source flux.  
To explore whether a closer background measurement might remove this 
discrepancy we repeated the aperture photometry measurements on the MIPS
24~$\mu$m observations using a 6\arcsec\ radius source aperture and a 
background aperture between 6\arcsec\ and 12\arcsec\ radius (which covers the
first Airy ring in the MIPS 24~$\mu$m PSF).  The aperture correction for
such measurements, based on our analysis of bright isolated sources, is 2.114.
We find that aperture photometry performed with a closer sky annulus
decreased, but did not eliminate, the discrepancy with the SAGE DR1 results,
and furthermore new discrepancies arose where previously there were none.
We conclude that our measurement at 24~$\mu$m may sometimes be overestimated 
for sources amid strong diffuse 24~$\mu$m emission.

\section{Identification of Candidate YSOs in the LMC}\label{ysocand}

YSOs are still enshrouded in dust that absorbs the stellar radiation
and irradiate at IR wavelengths; thus, they can be identified by
observed IR excesses.  Theoretical predictions of YSOs in
color-magnitude diagrams (CMDs) and color-color diagrams (CCDs)
are usually used to compare with observations in order to assess
the evolutionary stages of YSOs.  Low- and intermediate-mass YSOs
are known to be surrounded by circumstellar accretion disks and
envelopes, and their evolution is well established; thus, models
of their SEDs can be generated and predictions of their locations
in CMDs and CCDs can be made (e.g., Allen et al. 2004).  However,
the circumstellar structure and evolution of massive YSOs are much
less certain \citep[e.g.,][]{ZY07,Cetal07}.  To identify massive YSOs, 
either YSO models of low-mass stars are adopted \citep[e.g.,][]{Jetal05} or 
YSO models for massive stars are generated with the assumption that 
massive YSOs have circumstellar envelope and disk structures similar 
to their low-mass counterparts \citep[e.g.,][]{Wetal04,Retal06}.  
The SAGE team has identified YSOs by using the latter massive YSO models 
to predict locations of YSOs in different combinations of CMDs, and 
using known evolved stars, planetary nebulae, and other contaminants 
to mark regions in the CMDs to avoid \citep{Wetal08}.

To carry out an independent search for YSOs, we adopt a totally
different approach.  We start by examining mid-IR CMDs and CCDs
for the LMC and SWIRE data, and decide to adopt the [8.0] vs.
[4.5]$-$[8.0] CMD as the starting point to separate YSOs from
foreground and background contaminants, in agreement with the
suggestion of \citet{Hetal06}.  We use this CMD to select
YSO candidates first, then examined more closely each of the
candidate and considered  the source morphology, environment,
and spectral energy distribution (SED) over as wide a wavelength
range as possible to better assess their nature.

\subsection{Methodology: Minimizing Contaminants}

\begin{figure}
\plotone{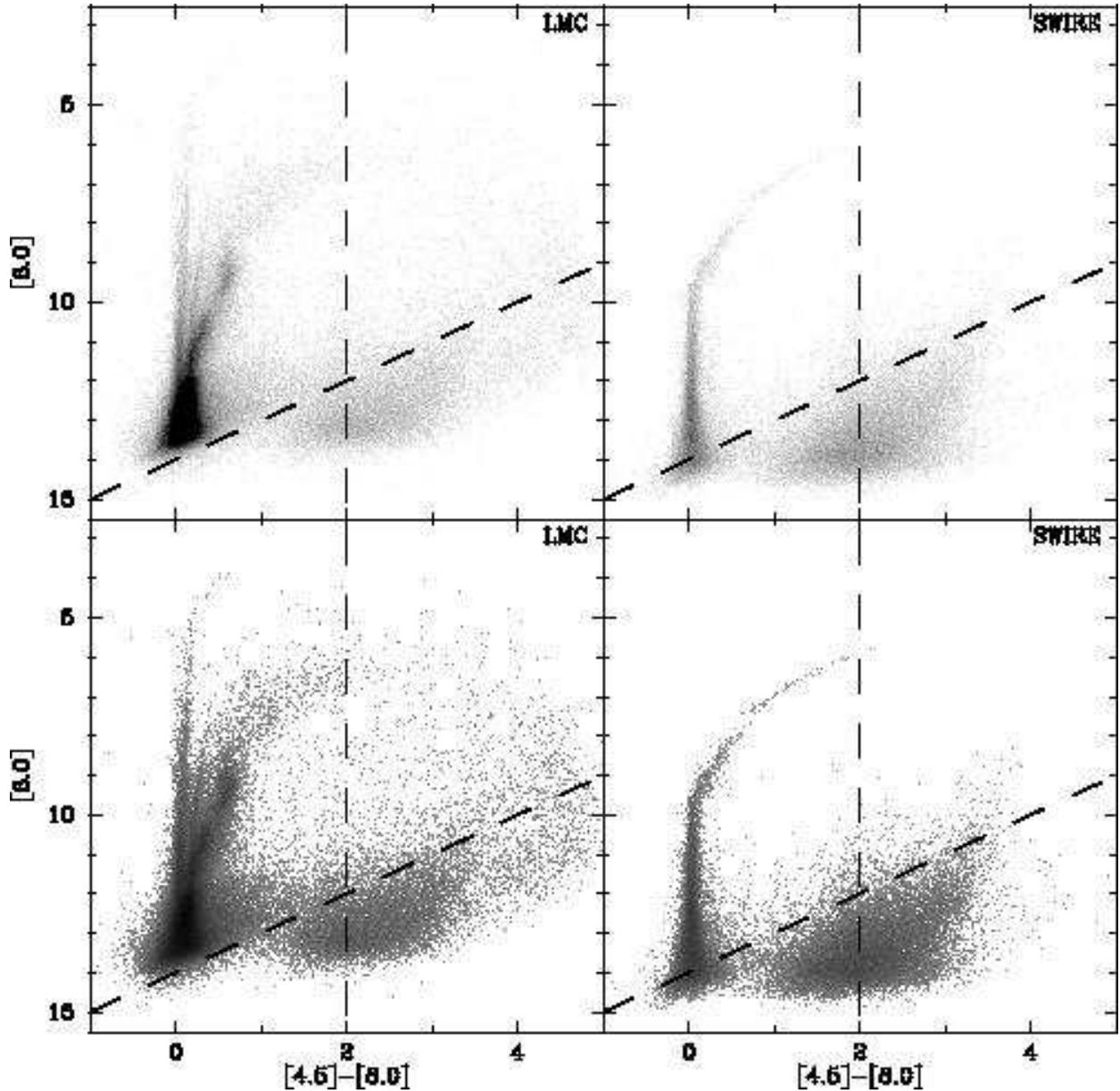}
\caption{Color-magnitude diagrams using IRAC [8.0] vs. [4.5]$-$[8.0] for
the LMC (left panels) and the SWIRE survey (right panels) shown as Hess
diagrams.  The panels in the top row use a square-root stretch while those
in the bottom row use a logarithmic stretch to make both sharp features
and the broad range covered by sources clear.  The vertical dashed line marks
the cutoff used to exclude most AGB, post-AGB, and main sequence stars 
(predominantly to the left), while the diaganol dashed line marks the
criteria used to exclude background galaxies (below this cutoff).  
Note that points with $[8.0] \gtrsim 5.0$ in the LMC survey and 
$[8.0] \gtrsim 9.0$ in the SWIRE survey are saturated.}
\label{CMD4-24_fig}
\end{figure}

In Figure~\ref{CMD4-24_fig} we present CMDs showing the IRAC 8.0~$\mu$m flux
with respect to the [4.5]$-$[8.0] color for all sources in the LMC and SWIRE
fields as Hess diagrams.  In these CMDs, main sequence stars will appear near
[4.5]$-$[8.0]$\simeq$0 while stars on the red giant branch (RGB) and 
asymptotic giant branch (AGB) form the features which branch to the 
right of the main sequence due to excess mid-IR emission arising from the 
dust in their stellar winds.  Examples of known LMC objects in similar
CMDs have been presented for massive stars \citep{Betal09} and AGB stars
\citep{Betal06}.  In the [8.0] vs. [4.5]$-$[8.0] CMD, YSOs are also expected 
to lie to the right of the main sequence as they are surrounded to varying 
degrees by the dusty remains of the protostellar core from which they formed
\citep[e.g.,][]{Aetal04,Retal06}.

To better understand the area over which AGB stars might be present in this
CMD we consider their expected colors and luminosities based on models for 
Galactic C- and O-rich AGB stars by \citet{G06}.  
While the LMC has a lower metallicity, AGB atmospheric models depend
on the stellar synthesized C and O abundances more than the initial 
abundances, thus the Galactic models are adequate in providing a
rough range of colors for the LMC AGB stars.
These models predict that most AGB stars have [4.5]$-$[8.0]$<$2 but 
that some rarer deeply enshrouded AGB stars redder than [4.5]$-$[8.0]$=$2
should be expected.  Thus, we have used [4.5]$-$[8.0]$>$2.0 as 
the first criterion to exclude AGB stars, main-sequence stars, and 
giants from our sample.  Recent analysis of the SAGE data by \citet{Betal06} 
have confirmed that this criterion will likely exclude most C and O-rich AGB 
stars but that some ``extreme'' AGB stars may be present in the sample.
The criteria of [4.5]$-$[8.0]$>$2.0 will exclude more
evolved YSOs that have dispersed most of their circumstellar dust, but we
are undertaking a follow-up program to identify these evolved massive YSOs.

To illustrate the locations of background galaxies in the CMD, we use
observations from the SWIRE Legacy Program, which was designed to 
study mid-IR properties of extragalactic sources.  The six fields 
observed in the SWIRE survey are all at high Galactic latitudes;
therefore, the two populations of objects that dominate the
[8.0] vs. [4.5]$-$[8.0] CMD for 
the SWIRE fields will be foreground Galactic stars and background galaxies.  
The SWIRE CMDs in Figure~\ref{CMD4-24_fig} shows that main-sequence foreground 
stars at [4.5]$-$[8.0]$\simeq$0 curve toward the red for [8.0]$<$9~mag 
where the sources begin to saturate due to the 30~s frame time used for these 
observations.  The remaining sources are dominated by distant galaxies.  
In order to exclude as many distant galaxies from our LMC data as possible 
but still retain as many possible YSO candidates, we use a color-magnitude cut 
where sources with [8.0]$>$14$-$([4.5]$-$[8.0]) are excluded \citep{Hetal06}.

\begin{figure}
\plotone{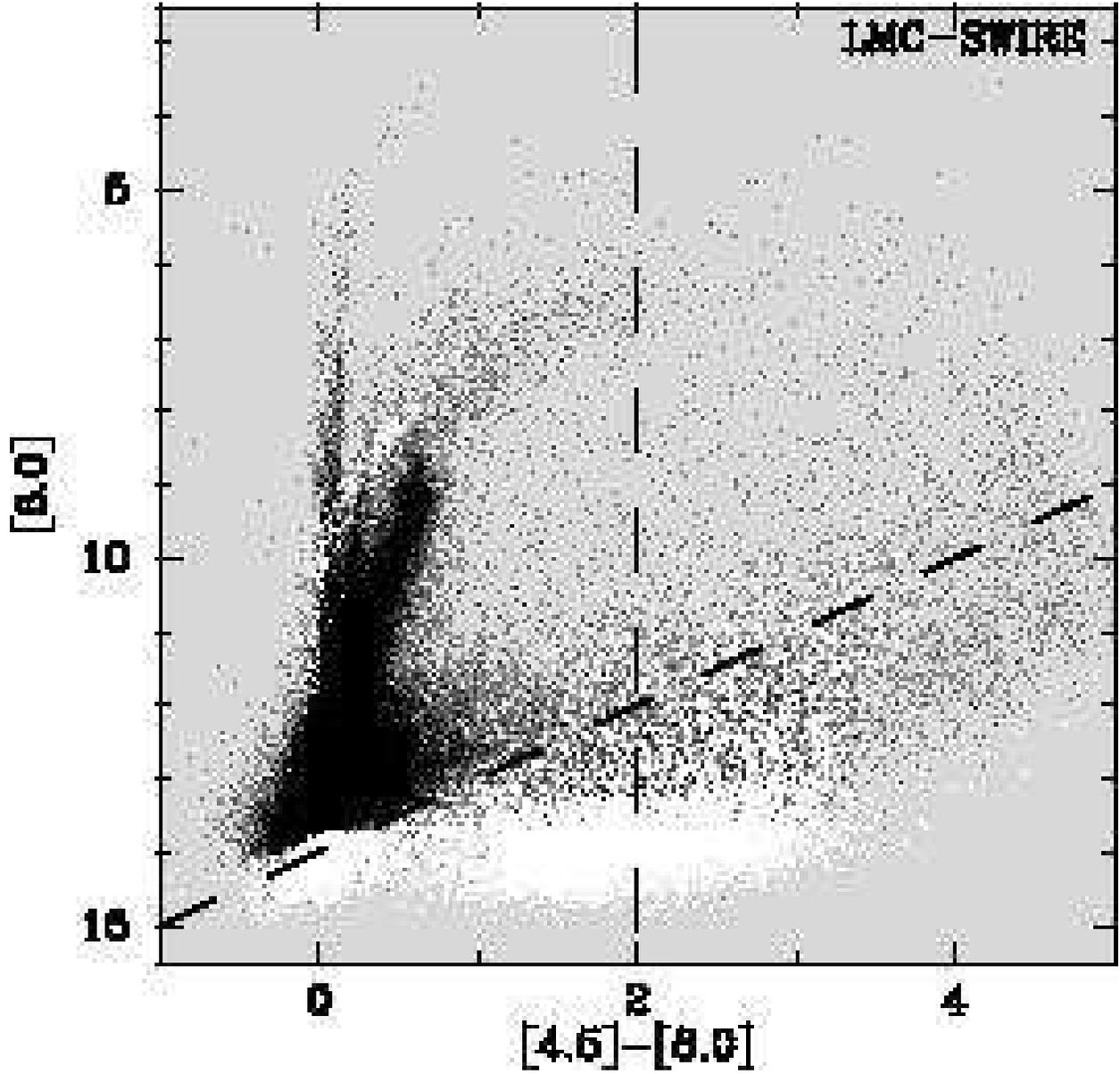}
\caption{The difference between the LMC and SWIRE color-magnitude Hess diagrams 
for the [8.0] vs. [4.5]$-$[8.0] CMD after normalizing by the survey areas.}
\label{LMC_diff_CMD}
\end{figure}

Within the wedge in the [8.0] vs. [4.5]$-$[8.0] CMD of the LMC defined by these 
two criteria we find 2910 sources remain.  As the total extinction through the LMC 
is less than that for Galactic molecular clouds, from which the empirical criterion
was derived \citep{Hetal06}, we expect some background galaxies may be present 
in our sample.  To estimate the extent to which background galaxies should be 
present we have examined the same wedge in the SWIRE CMD.  After eliminating
sources from the saturated main sequence we find 859 sources remain.  We then use the 
4.5 and 8.0~$\mu$m images from the SWIRE and LMC surveys to estimate the 
area covered and find that the total area covered by SWIRE is $\sim$48.2 
square degrees while the LMC survey covers 61.7 square degrees.  Assuming 
that both surveys are complete in this region of the CMD we estimate that 
$\sim$1100 background galaxies should still contaminate our initial set of 
candidates.
Furthermore, since the SWIRE Survey data are deeper than the LMC survey 
data we can test whether a sample of candidate YSOs selected from the 
sources in this region of the CMD is affected by the completeness limits 
of our photometry.  To do this we normalize the Hess diagrams by the area 
covered by the survey from which they are taken and subtract the SWIRE CMD 
from the LMC CMD.  The resulting differenced Hess diagram is shown in 
Figure~\ref{LMC_diff_CMD}.  The areas in Figure~\ref{LMC_diff_CMD} where the 
deeper SWIRE observations consistently have more source counts appear white 
and demonstrate that the area of the [8.0] vs. [4.5]$-$[8.0] CMD we are 
using to search for YSOs should be complete excluding regions with 
exceptional crowding or with a complex or bright background (see Figure~\ref{comp_YSO}).

\subsection{Methodology: Verification of YSO Candidates and Further Elimination 
of Contaminating Sources}

\subsubsection{Additional Supporting Observations}

While the 2910 sources thus selected could be YSOs, among them there should still 
be a significant number of AGB stars, post-AGB stars, and galaxies.  The LMC 
has been surveyed at many wavelengths and we will now combine the results 
from some of these other surveys to better assess the nature of the selected 
sources using both imaging and photometric results.  We use images to 
assess not only the source morphology but also the surrounding interstellar
environment.  At the same time we use existing photometric surveys to 
extend the SED from the mid-IR to the near-IR and optical wavelengths.

Specifically, we have downloaded the red images from the Digitized Sky 
Survey (DSS2r), obtained the H$\alpha$ images\footnote{The MCELS H$\alpha$ images 
are not continuum subtracted.} from the Magellanic Clouds 
Emission Line Survey \citep[MCELS;][]{Setal99}, and downloaded 
the $J$, $H$, and $K_s$-band images from 2MASS \citep{Setal06} for comparison 
with the IRAC and MIPS images.  
To extend the SEDs of sources we have used the near-IR photometry from 
the 2MASS PSC \citep{Setal06} and the optical photometry 
from the Magellanic Clouds Photometric Survey \citep[MCPS;][]{Zetal04}.  
The 2MASS PSC has completeness limits in the absence of confusion of 15.8, 
15.1 and 14.3 mag (or 0.78, 0.97, and 1.3~mJy) in the $J$, $H$, and $K_s$-bands,
respectively.  In the MCPS survey there is little evidence for incompleteness 
at $V <$~20~mag ($<$36~$\mu$Jy), but the survey shows severe incompleteness
at $UBVI$ of 21.5, 23.5, 23, and 22~mag (4.3, 1.8, 2.3, and 3.8~$\mu$Jy), 
respectively.
In order to identify potential counterparts among the near-IR and optical 
catalogs, we cross-correlated the positions of the sources from 2MASS and 
MCPS with the positions of sources from our IRAC and MIPS photometry, and
a positive match was considered to be closer than 1\farcs0.  In the rare
cases where an IRAC source has multiple  matches in the 2MASS or MCPS catalogs, 
the closest near-IR/optical source is assumed to be the counterpart.

For many regions in the LMC we have obtained observations using the MOSAIC
and ISPI cameras on the CTIO Blanco 4m telescope (see \S\ref{sec_obsoptnir}).
These images have higher angular resolution (typically sub-arcsecond) and 
better sensitivity than the DSS2r or 2MASS images.  
Where possible we use these MOSAIC $I$-band and 
ISPI $J$ and $K_s$-band images to further supplement our analysis as they 
are better able to reveal fainter background galaxies, detect faint 
optical and near-IR counterparts to sources, and resolve groups and 
clusters associated with the YSO candidates.

To obtain a rough assessment of the molecular gas environment we use the 
available maps from the NANTEN CO(J=1-0) survey of the LMC \citep{Fetal01}
which covers most of the area observed by {\it Spitzer}.
These maps allow us to determine whether or not each source is 
associated with giant molecular clouds (GMCs).  The NANTEN survey is 
sensitive to GMCs with masses greater than $\sim$1-2$\times10^{4}$~M$_\odot$.

\subsubsection{Multi-Wavelength Assessment of Sources}

\begin{figure}
\centering
\includegraphics[width=3.8in]{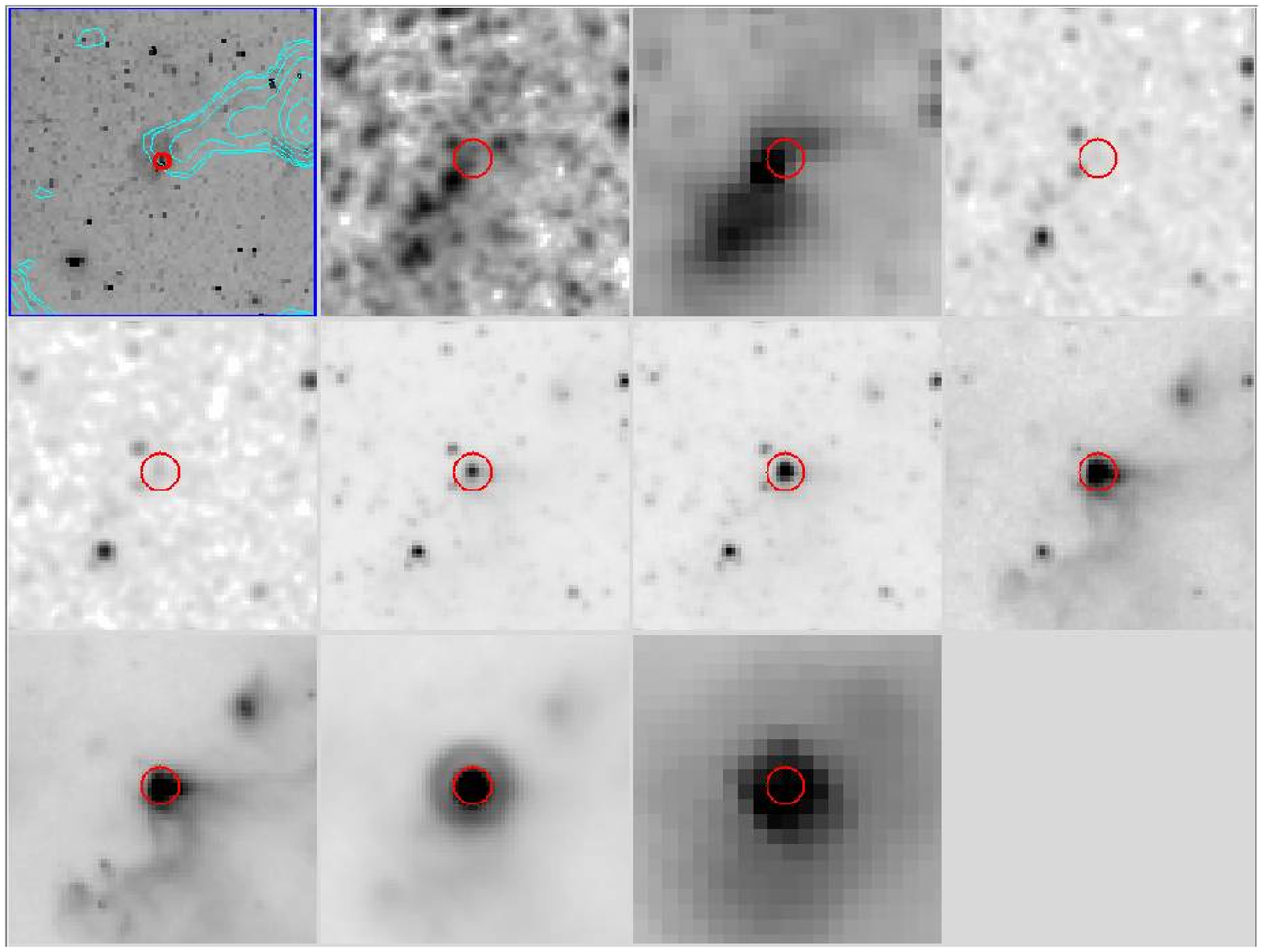}
\hfil
\includegraphics[width=2.2in]{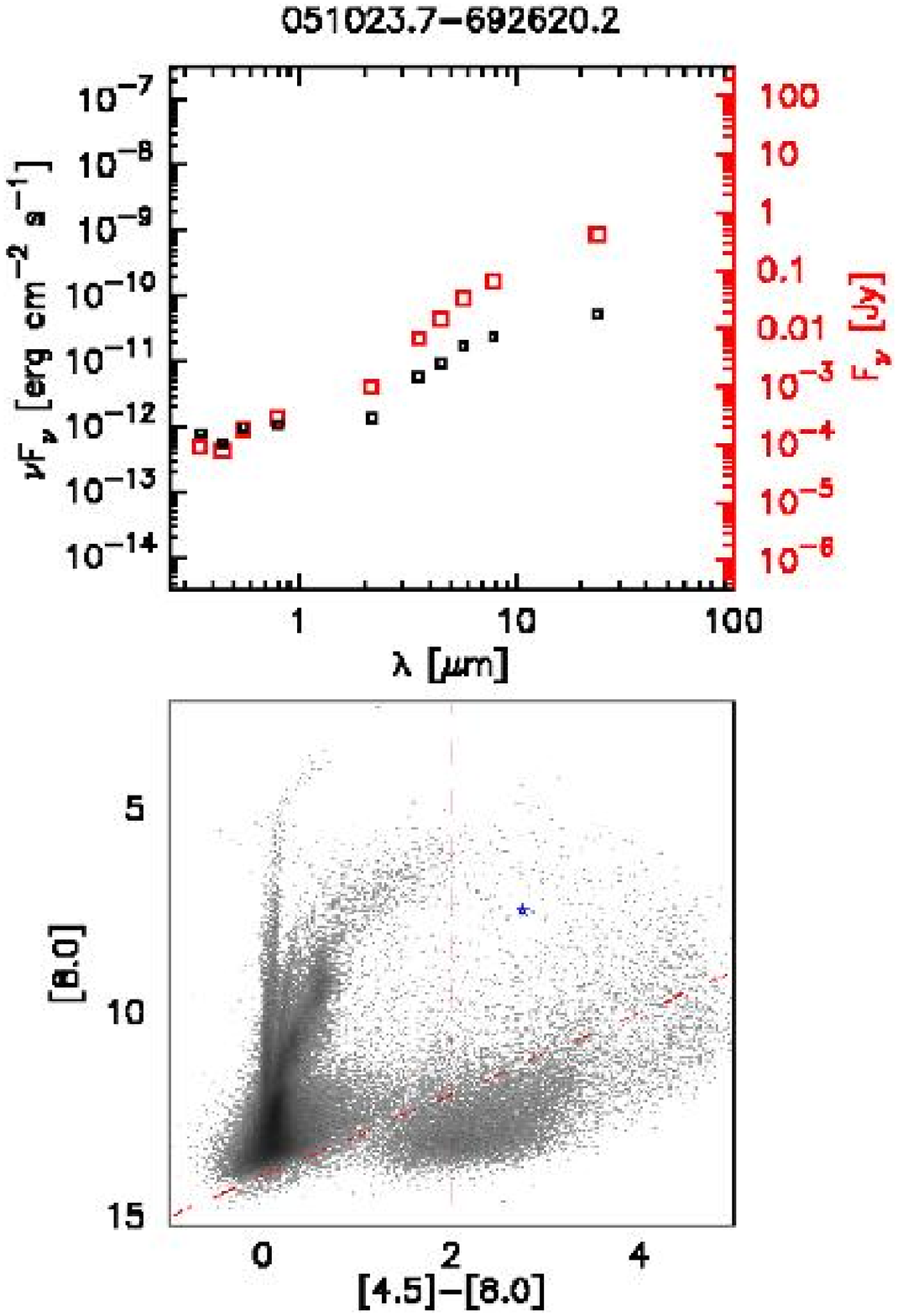}
\caption{A screen capture showing images and plots used when classifying each
candidate. The images on the left are for the source 051023.7$-$692620.2.
The image in the top-left corner has a wide field to show the location within the 
LMC using an MCELS-\ha+continuum image overlaid with CO contours from the NANTEN survey.  
The remaining images show a close-up view of the source.  From top-to-bottom and 
left-to-right those images are: DSS2r, MCELS-\ha+continuum, 2MASS-J, 
2MASS-K, IRAC 3.6, IRAC 4.5, IRAC 5.8, IRAC 8.0, MIPS 24, MIPS 70.  In this screen 
capture the wide-field image has a $\sim$20\arcmin\ extent while the other panels show an 
$\sim$80\arcsec$\times$80\arcsec\ field.  The solid circle
in the center of each image has a 5\arcsec\ radius.  The plots
on the right show the spectral energy distribution for the source both as 
$\nu$F$_\nu$ vs.\ wavelength and F$_\nu$ vs.\ wavelength as well as the source
location marked as a blue star in the [8.0] vs. [4.5]-[8.0] CMD.  The DS9 utility 
allows changes in image contrast, magnification, and a cross-check for the alignment
of the source at different wavelengths.  (A color version of this figure
is available in the electronic edition only.)}
\label{fig_screen}
\notetoeditor{This figure should appear in color in the electronic edition of the Journal only.}
\end{figure}

We used this large volume of multi-wavelength observations extending
from optical to mid-IR wavelengths to assess
the nature of each source in our initial sample.  To this end we have 
constructed a library of ``postage stamp" images with a field-of-view 
$\sim$5\arcmin$\times$5\arcmin\ ($\sim$75$\times$75~pc) centered at the 
location of the IRAC source for wavelengths ranging from $\sim$6000~\AA\ to
70~$\mu$m, specifically: red continuum, H$\alpha$, $J$-band, $K_s$-band, 
3.6~$\mu$m, 4.5~$\mu$m, 5.8~$\mu$m, 8.0~$\mu$m, 24~$\mu$m, and 70~$\mu$m
images.  
Using the flexible capabilities of DS9 \citep{JM03} we then simultaneously 
displayed and considered these postage-stamp images along with the SED
based on the photometry to assess the nature of the source.  In doing
so we consider not only the relative source brightness in multiple 
wavelengths but also the source morphology,  the immediate stellar 
and interstellar environment, and the nature of other sources in that 
environment.  Figure~\ref{fig_screen} shows an example for one candidate
of the graphical information displayed by the software used when classifying
sources.  We further used the SIMBAD Astronomical Database to search
for previously identified sources and considered any previous assessment 
of their nature.

\section{Results}

The multi-wavelength assessment of each source was made independently by 
both authors and the results were compared to reach a final consensus.  
The resulting taxonomy splits the candidate sources into five categories: 
(1) evolved stars, (2) planetary nebulae, (3) background
galaxies, (4) diffuse sources, and (5) definite, probable, and possible YSO 
candidates.  Below we describe the characteristics common to most objects
in each category, present a few illustrative examples, and provide
tabulated lists of sources with their photometric measurements.  In the
following description of the common characteristics of each type of source,
it should be stressed that the nature of each source is never assessed based
on one single characteristic, instead it is based on a consensus of all available
criteria.
In many cases, particularly for fainter sources, the results of this analysis 
do not reach a definitive assessment as to the nature or classification of a
source but instead reach a conclusion that is better described as a 
likelihood for the nature of a source.

\subsection{AGB and post-AGB stars}

In \S4.1 we chose a mid-IR color criteria that would exclude most
AGB and post-AGB stars based on the models of \citet{G06}.  These models
suggest that there should still exist some extreme objects which are 
deeply embedded in a dusty envelope that will have mid-IR colors 
and brightnesses consistent with our preliminary YSO source criteria.
These sources can be identified when considering their 
images and SEDs because (1) they are typically not associated with any
interstellar gas or dust structures, (2) their mid-IR SEDs peak in the 
IRAC bands ($<$8.0~$\mu$m) and drop in the MIPS 24~$\mu$m and 70~$\mu$m 
bands, and (3) their mid-IR SEDs appear roughly consistent with 
that of a blackbody with temperature between $\sim$400 and 1000~K.  
In some cases these sources show an SED with two peaks one in the 
optical/near-IR and one in the mid-IR similar to the SEDs of 
post-AGB stars with dusty tori \citep[e.g.,][]{deRetal06}.

\begin{figure}
\plotone{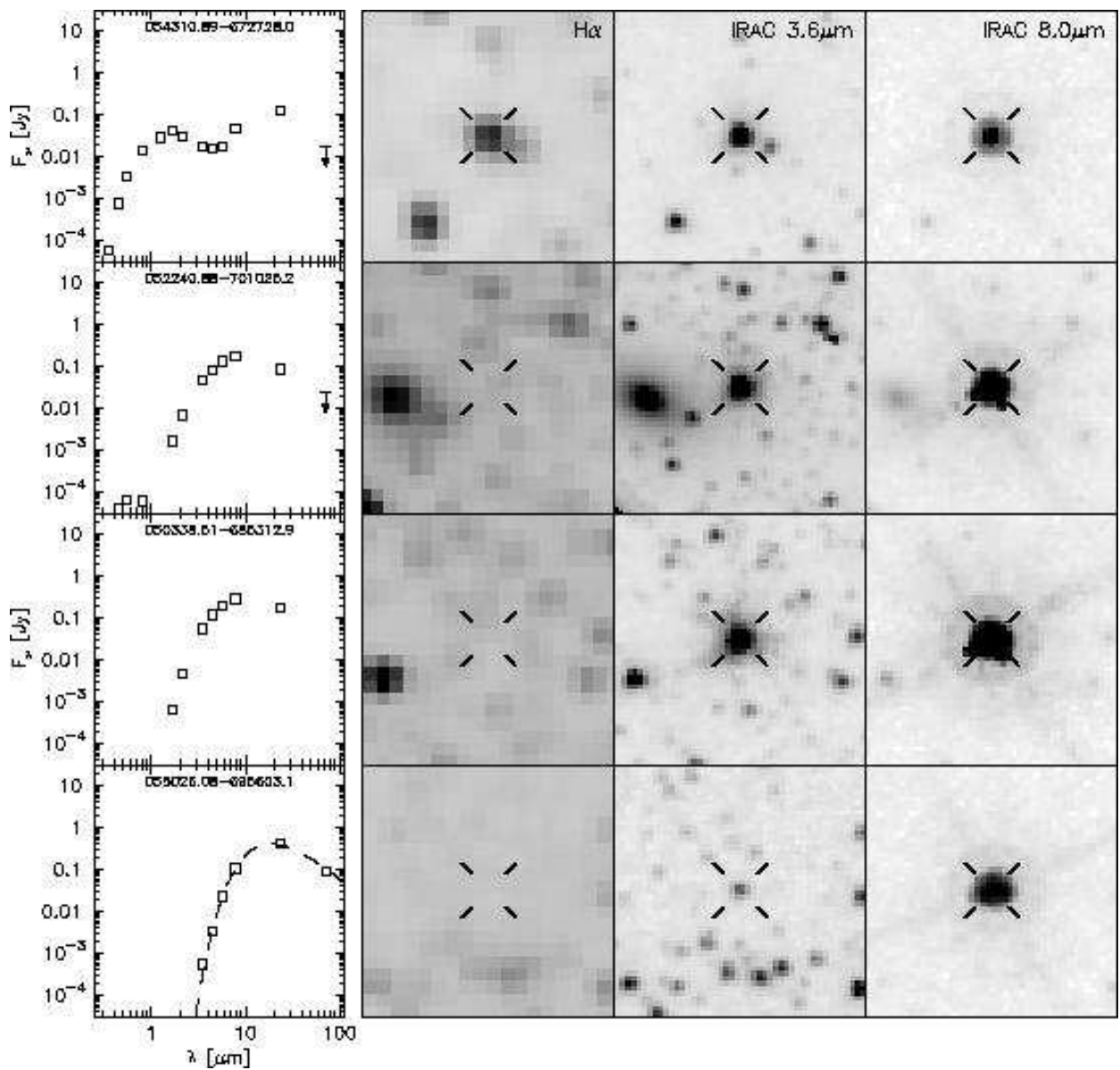}
\caption{Examples for sources that have been identified as AGB/post-ABG stars.  
The panels in the left column show the SEDs for each source.  The optical
near-IR, and mid-IR photometry are from the MCPS, 2MASS, and this paper,
respectively.  Three-$\sigma$ upper limits are indicated by an arrow where
applicable.  To the right of each SED are images of the source at H$\alpha+$continuum  
(MCELS), IRAC 3.6~$\mu$m and IRAC 8.0~$\mu$m.  The field of view for each 
image is 60\arcsec$\times$60\arcsec .  The SED and images in the bottom row
are for a source identified as an ERO which has no optical or near-IR counterpart.
The best-fit single temperature black-body for this ERO with T=274$\pm$55~K is 
shown as a dashed line.}
\label{fig_exampleAGB}
\end{figure}

We find that roughly 111 of the 442 sources with [8.0]$<$8.0~mag to be
likely either AGB or post-AGB stars, but among sources with
[8.0]$>$8.0~mag only six sources that fit the same description 
are present.  These sources appear similar to the ``extreme'' AGB stars
noted by \citet{Betal06}.  In Table~\ref{tab_evolvedstar} we summarize the 
positions, mid-IR photometric measurements, and cross-identifications with 
known sources using common designations from the SIMBAD database for the 
sources we have classified as AGB and post-AGB stars.  We show four 
examples of sources from this category in Figure~\ref{fig_exampleAGB}.

In addition to the ``normal" AGB and post-AGB stars we find 13 sources with 
[4.5]-[8.0]$>$4~mag with no optical or near-IR 
counterpart in the DSS or 2MASS surveys.  Eleven of the 13 sources have
[8.0]$<$8.0 mag.  We refer to these sources
as Extremely Red Objects (EROs) as their mid-IR colors are much redder
than the typical sources we identify as AGB and post-AGB stars.  Their 
SEDs appear to peak between 8 and 24~$\mu$m and have either much 
lower 70~$\mu$m fluxes or are not significantly detected in the 
MIPS~70~$\mu$m observations.
All 13 of these sources are in isolated environments, not associated
with interstellar gas or dust structures.  They do not appear to be 
concentrated toward the LMC Bar nor does their spatial distribution
noticeably match that of the underlying LMC stellar distribution.  
These objects appear to form a unique class, although some of them have been
suggested to be evolved stars or YSOs \citep{Letal97,Wetal08}.
Subsequent IRS observations of 7 of these EROs have revealed that they
are extreme carbon stars \citep{Getal08}.  Therefore we include these
sources among those in Table~\ref{tab_evolvedstar} where their class is
designated ``E."  The bottom panel in Figure~\ref{fig_exampleAGB} 
shows an example of one or these sources.

\subsection{Planetary Nebulae}

\begin{figure}
\plotone{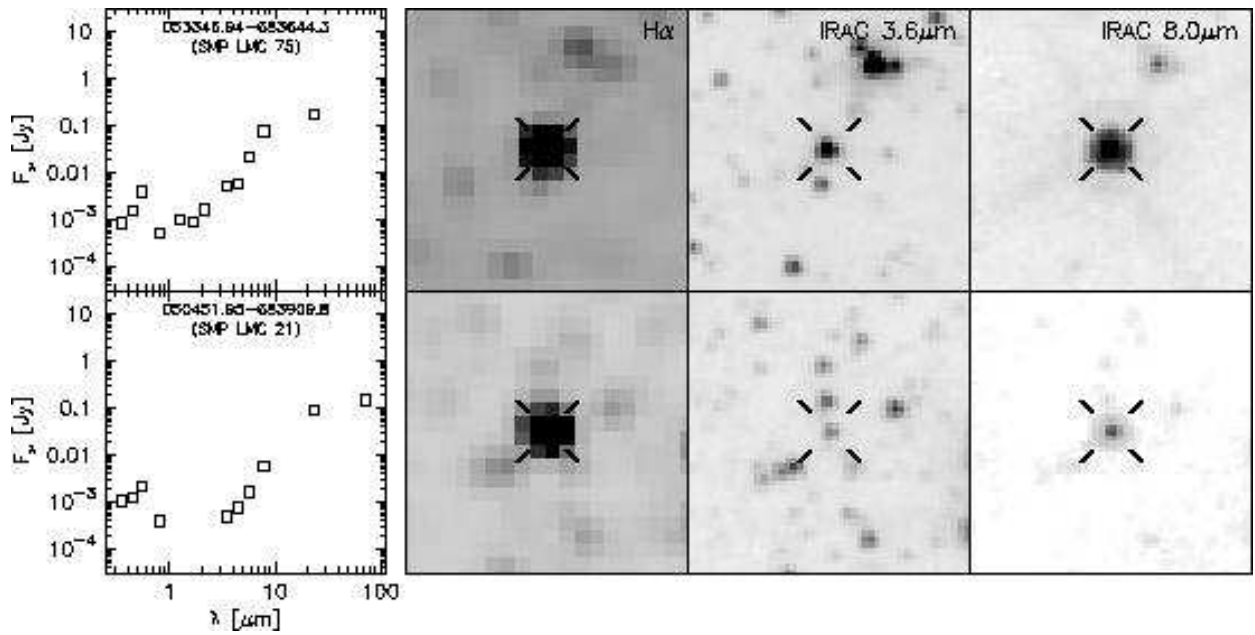}
\caption{Examples for sources that have been identified as PNe.
The panels in the left column show the SEDs for each source with images from
the MCELS-H$\alpha$+continuum, IRAC 3.6~$\mu$m and IRAC 8.0~$\mu$m shown to the right.}
\label{fig_examplePN}
\end{figure}

We find that 53 of the initial sample are coincident with objects that
have been previously identified as likely LMC planetary nebulae 
\citep[e.g.,][]{LM63,SMP78,MGPN92,RP06}.
These sources are primarily identified through our
search for known counterparts but are also evident as they are generally
marginally resolved or unresolved H$\alpha$ sources in the MCELS survey, and
further they typically have SEDs that appear irregular, reflecting that the 
primary emission at different wavelengths alternate among nebular lines, 
PAH, and dust emission.  Two examples are
shown in Figures~\ref{fig_examplePN} and the positions, photometric 
measurements, and cross-identifications to known sources are summarized 
in Table~\ref{tab_pne}.

\subsection{Background Galaxies}\label{sec_bkggal}

Based on comparisons with results from the SWIRE data we estimate roughly 
$\sim$1,100 background galaxies should be present among our initial sample of 
sources, and we expect this population to be evenly distributed 
throughout the survey area.  We are able to identify these galaxies through 
three complementary methods.

\begin{figure}
\plotone{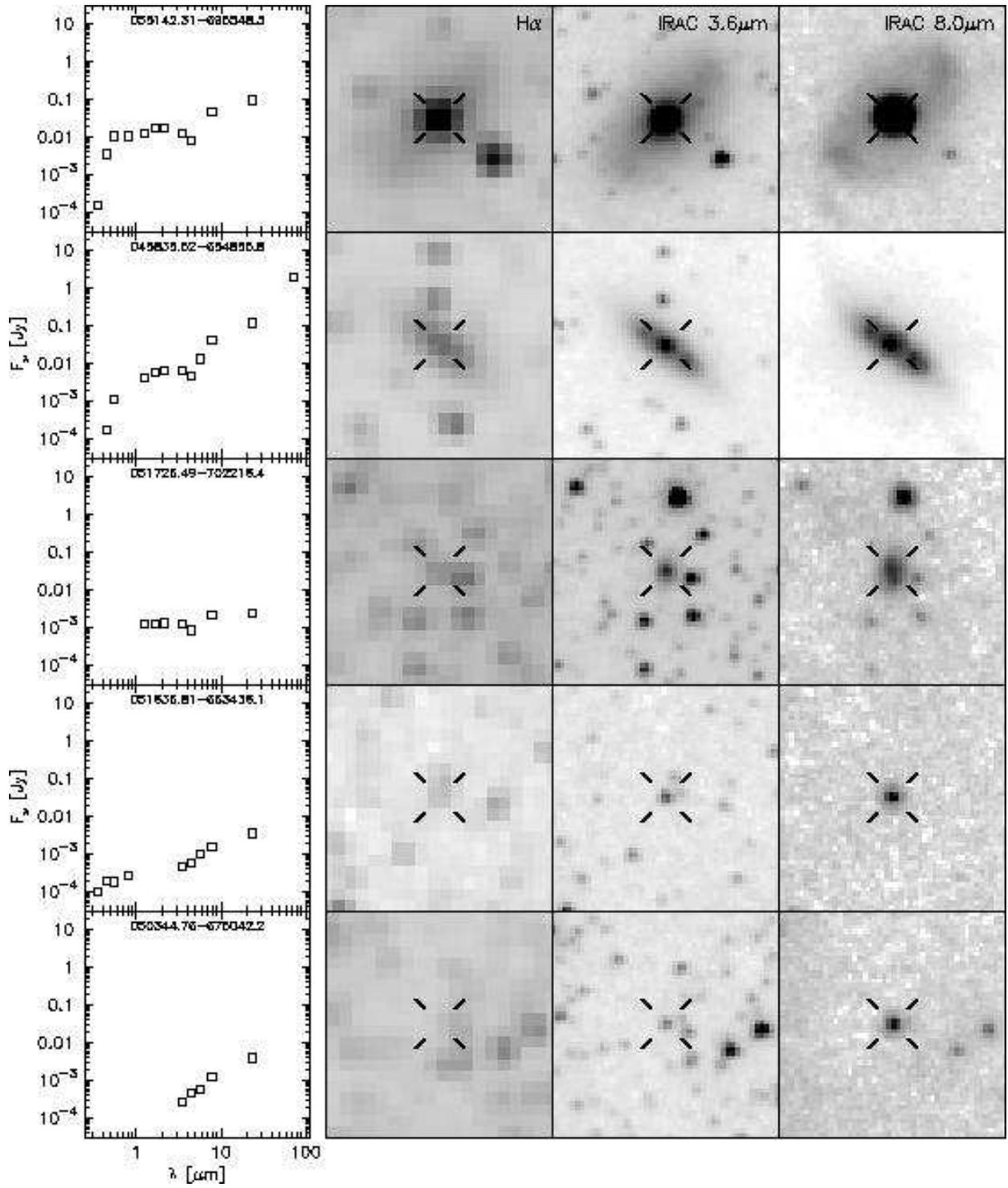}
\caption{Similar to Figure~\ref{fig_exampleAGB} but for sources that have been 
identified as galaxies.}
\label{fig_exampleGAL}
\end{figure}

First, a large fraction of the background galaxies are easily identified by 
closely inspecting the images as they are extended sources 
(see Figure~\ref{fig_exampleGAL}).  In these cases the galaxy has been 
identified as a potential candidate in our catalog of point sources because 
either the galaxy is only marginally resolved or the unresolved nuclear region 
has been identified as a point source.  Second, by examining the SED that 
results from these resolved and marginally resolved galaxies\footnote{
These galaxies are resolved at some or all wavelengths but the aperture 
photometry was performed assuming a point source.  The observed SED for the 
background galaxies reported in this paper should not be treated as though 
it were the true integrated SED of a galaxy.} 
we are able to further identify some fainter and unresolved background 
galaxies with similar SEDs as is the case of the last two examples in 
Figure~\ref{fig_exampleGAL}.  
Finally, there exist a set of sources with flat or slightly rising
SEDs.  We find that these SEDs are similar to those of known background
Seyfert galaxies and quasi-stellar objects in our sample.  
Based on the similarity in SEDs and their being isolated from 
interstellar gas and dust structures, we tentatively identify these 
sources as background galaxies.

We have divided the background galaxies into two subsets that reflect 
the relative certainty of the sources being background galaxies.  
The first subset, ``background galaxies," are those sources which are 
clearly extended or which appear to be marginally resolved and have an 
SED that matches those of the extended sources.  The second subset, 
``probable galaxies," do not appear extended, are generally isolated 
(not associated with any interstellar structures), and have 
SEDs that are similar to those of either resolved galaxies or 
known active galaxies.  We find that the range of SEDs seen among
those categorized as ``background galaxies" and ``probable
galaxies" are similar to those seen in other samples of distant galaxies
observed by {\it Spitzer} \citep[e.g.,][]{Detal08}.
The positions, photometric measurements and cross-identifications with 
previously known objects are summarized for these sources 
in Tables~\ref{tab_gal} and \ref{tab_pgal}.  
We find 959 and 116 sources that are categorized as 
background galaxies and probable background galaxies, respectively.  As 
this matches the prediction of 1,100 background galaxies to within a few
percent we expect that the contamination by unidentified background 
galaxies in our final sample of YSO candidates should be minimal.
We caution that this should not be interpreted to mean that the 
classification of any individual object as a background galaxy is correct.

\subsection{Diffuse Sources}\label{sec_diff}

We have found a moderate contamination among our initial sample 
by ``sources'' that are local enhancements within filamentary dust 
emission most notably at either intersections between filaments or at sharp 
bends in filaments.  We refer to these as ``diffuse sources.''  Diffuse 
sources are identified through careful inspection of the images, 
particularly through comparison of the IRAC images where the sources 
were identified, and searches for counterparts in the near-IR and 
MIPS images.  When such a feature appears to be an enhancement in a 
filament and no evidence for a counterpart to the local enhancement 
in the IRAC bands is seen at longer or shorter wavelengths, we place 
the source in the diffuse category.  In searching for near-IR counterparts 
to an IRAC source, the ISPI $J$ and $K_s$-band images, which are typically 
2-3~mag deeper than the 2MASS Survey have proven to be powerful tools as they 
place a strong limit to the existence of a near-IR counterpart. 

\begin{figure}
\plotone{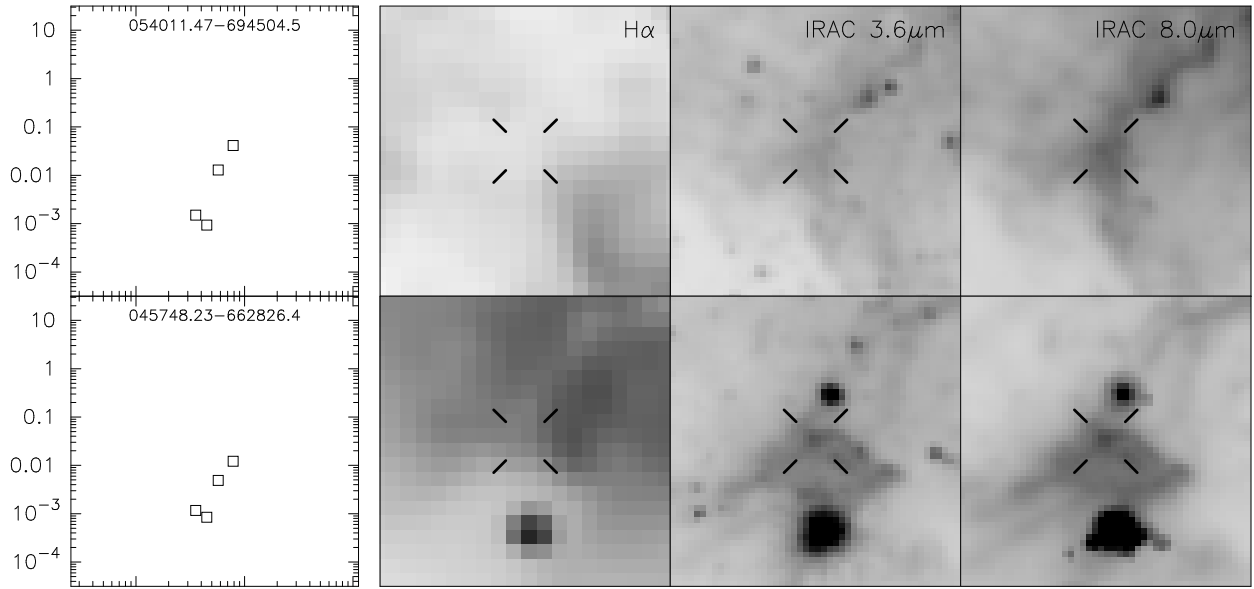}
\caption{Similar to Figure~\ref{fig_exampleAGB} but for candidates that 
have are diffuse ``sources'' (see \S~\ref{sec_diff} for an explanation).}
\label{fig_exampleDIFF}
\end{figure}

A total of 159 sources were classified as diffuse.
In the examples shown in Figure~\ref{fig_exampleDIFF} we see that these 
``sources" typically exhibit a mid-IR SED indicative of PAH emission 
\citep[e.g.,][]{Jetal05,Getal04}.  Although these are most likely 
local peaks in the interstellar dust emission rather than true point 
sources, we summarize their locations and properties in Table~\ref{tab_diff}.
Note that these diffuse sources might contain low- or even intermediate-mass
YSOs as {\it HST} NICMOS images of a few diffuse sources in the 30\,Dor region
have revealed Herbig Ae/Be stars and lower mass pre-main-sequence stars
\citep{Betal01}.

\subsection{Young Stellar Objects}

Both images and the SED of each source are critical in the identification
of YSOs as these sources can have distinctly different properties depending
on their evolutionary state and their environment.  Due to their dusty 
surroundings we expect YSOs to have significant mid-IR emission, but this
emission and the SED at other wavelengths will vary depending on the 
distribution and geometry of any circumstellar disk and envelope material.
Therefore, there is not a single set of criteria that describe all YSO
candidates but rather a broad range that can encompass a wide variety of
sources.

The YSO sources we have identified have SEDs that rise in the mid-IR
between 3.6 and 8.0~$\mu$m, which is not surprising due to the initial
criteria that [4.5]$-$[8.0]$>$2.0.  Furthermore, most YSO candidates  
have counterparts in the MIPS 24~$\mu$m images (note that in some cases no
photometric measurement is available at 24~$\mu$m due to crowding,
saturation, or the presence of bright diffuse emission).  In many cases 
we find that YSO candidates have near-IR and sometimes even optical counterparts 
but these can reflect the evolutionary stages of the candidates or whether
there are deeper ISPI and MOSAIC images available.

We have divided the YSO candidates into three groups that reflect our confidence
as to their true nature: (1) ``Definite YSOs" where we are highly confident of their
nature,
(2) ``Probable YSOs" where some characteristic of the source points to 
a possible alternative nature, such as a background galaxy, and 
(3) ``Possible YSOs" where we have a higher confidence that the source 
is not a YSO but cannot definitively rule out that the source is a YSO.  
The third group of sources, possible YSOs, are included for completeness 
so that these sources are not summarily dismissed from further consideration 
of their true nature.

In Figures~13--\ref{fig_exampleYSO3} we show examples from 
all three groups of sources.  Furthermore, Figures~13b-–x and
Figures~14b\&c present SEDs and images for all 247 sources 
classified as definite or probable YSOs with [8.0]$<$8 and are available in the
electronic edition of the Journal. 
In general, sources in our first two groups, definite and probable
YSOs, have a clear 24~$\mu$m counterpart, and the brighter sources also have a 
70~$\mu$m counterpart.  
The sources in the third group, possible YSOs, tend to be among the fainter 
sources where 24~$\mu$m and 70~$\mu$m counterparts could lie below the
detection threshold or are more easily hidden amid diffuse emission.  
Among the original 2910 candidates 855 have been classified as definite YSOs, 317 as
probable YSOs, and 213 as possible YSOs.
The positions and photometric measurements of the definite, probable, and
possible YSOs are summarized in Tables~\ref{tab_yso1}--\ref{tab_yso3}, 
respectively.

\subsection{Resulting Classification System}

Using the above broad categories each source in our initial sample was assigned 
a class based on each author's assessment.  For the brightest sources (those
with [8.0]$<$8), both authors classified each source on two different occasions.
For the fainter sources, the first author went through the classification process
for all sources twice.  These results were then intercompared and 
synthesized to obtain a final classification for each source.  The classes for 
AGB/post-AGB stars, EROs, PNe, background galaxies, diffuse sources, and candidate 
YSOs were respectively designated with A, E, P, G, D, and C.  In order to 
accommodate the degree of certainty that a source fit into a class, the 
option was given to suggest a secondary or alternate class for each source.  
For example, a source that was most likely a YSO candidate but had many 
characteristics that suggested its true nature might still be a background 
galaxy would be classified with the designation CG and would fall into the 
broad category of Probable YSO.  On the other hand, if the source was most 
likely a galaxy but a YSO candidate could not be ruled out the designation 
would have been GC and the source would be included in the both the probable
galaxy class and in the possible YSO class.  Consequently, such a source 
would appear in both the table of probable background galaxies 
(Table~\ref{tab_pgal}) and in the table of possible YSOs (Table~\ref{tab_yso3}).  
Figure~\ref{fig_exampleYSO2} shows sources classified as probable YSOs and 
has examples of sources with a CG classification while 
Figure~\ref{fig_exampleYSO3} shows sources classified as possible YSOs and 
has an example of a source in the GC class. 

In our analysis, two other broad classes were needed 
to accommodate a number of sources.  The first of these were ``normal" stars,
which include objects exhibiting normal stellar photospheric emission
and excess mid-IR emission due to circumstellar dust.  These sources were
designated with ``S."  The positions and photometric measurements of the 293
sources we have classified as stellar are summarized in Table~\ref{tab_stellar}.
The sources in Table~\ref{tab_stellar} are dominated by three classifications:
64 ``normal'' stars (class S), 165 stars with photometric contamination from 
diffuse emission (class SD), and 39 stars with an IR excess that could possibly 
be YSOs (class SC).  The second broad class consists of rejected sources.  
These are generally sources near the edge of the survey region where 
incomplete sets of observations prohibit proper classifications.  There are 
a few other cases where unrejected transients 
(e.g., cosmic ray hits) had resulted in faulty photometric measurements and hence 
source SED.  These sources were designated ``R" and are not among the tabulated sources.  

A follow-up {\it Spitzer} IRS program (PID: 40650) obtained observations of 269 of 
our brighter, definite and probable YSOs.  The first results from this program show 
that $\gtrsim$95\% of these objects have mid-IR spectra in the range between 
$\sim$5--37~$\mu$m consistent with our assessment that they are YSOs \citep{Setal09}.  
The YSOs candidates included in this IRS survey included most of those with 
[8.0]$<$8.0~mag but did not extend to sources fainter than [8.0]=9.0 mag.  
Therefore, we conclude that a majority of the brighter YSOs candidates reported here 
are indeed YSOs.

\section{Discussion}

Using the results from our multi-wavelength assessment of YSO candidates it 
is now possible to examine the observed properties of the sources with respect 
to the classification we have made.  
In Table~\ref{tab_summary_class} we summarize the number of objects that have
been assigned into each class described in the previous section.  In this
section we examine the spatial and photometric properties of the sources
and compare our results to those of the SAGE team.

\subsection{The Spatial Distribution of Red Mid-IR sources in the LMC}

Figures~\ref{fig_distOTHER} and \ref{fig_distYSO} show the location of 
sources with different classes throughout the LMC.  The AGB/post-AGB stars
and PNe populations will be incomplete but do appear more concentrated 
toward the LMC center as expected if they follow the underlying stellar 
distribution.  In contrast the background galaxies appear randomly distributed
across the field while the diffuse sources are concentrated in regions with
high H$\alpha$ surface brightness.  The distribution of background
galaxies is consistent with a homogeneous population of distant sources.  
Similarly, the diffuse sources tend to lie in and around \ion{H}{2} regions
consistent with our appraisal that these faux sources are knots in dusty
nebular material.  Furthermore, the relatively higher UV radiation field in 
the vicinity of the \ion{H}{2} regions provides a natural explanation for
why the SEDs often appear to be dominated by emission from PAHs.

In Figure~\ref{fig_distYSO} the sources we have classified as definite, 
probable, and possible YSOs show a completely different distribution.
They tend to be concentrated in or around either molecular clouds or
\ion{H}{2} regions.  Moreover, this tendency becomes more pronounced if 
the candidates classified as possible YSOs are excluded.  This implies 
that the YSOs are roughly correlated with the concentrations of dense gas 
or regions where active massive star formation took place within the last 
few Myr.  This latter association demonstrates that star formation rarely takes
place as a single event but rather is extended in the time domain.

We note that those sources classified as possible YSOs should generally 
be excluded when analyzing the overall star formation properties in the LMC.
This is in line with our description of the ``possible YSO'' source class 
as these are sources whose nature had a more likely alternate explanation but
where a YSO nature was not completely excluded.  We conclude that the 
distributions of YSO candidates along with the other source populations 
provide a general validation to our classification efforts.

\subsection{Using Color-Color and Color-Magnitude Criteria to Classify 
Sources}

We now use the source classifications to examine whether there are regions 
within color-color and color-magnitude space that can be combined to more
effectively discriminate YSOs from the other types of sources that tend to
share some photometric properties.  In 
Figure~\ref{fig_CMDbyclass} and Figure~\ref{fig_CCDbyclass} we present a 
variety of CMDs  and color-color diagrams (CCDs) for different classes of 
sources.  

In the cases of AGB/post-AGB stars and PNe the populations suffer from
small number statistics.  Based on the CMDs and CCDs in 
Figures~\ref{fig_CMDbyclass}\&\ref{fig_CCDbyclass} it appears that many of 
the AGB/post-AGB stars could be identified using a complex color criteria, but
that consideration of each source SEDs would still be necessary as a final 
discriminant to separate all AGB and post-AGB stars from the YSOs.  The PNe 
on the other hand appear to mix among the YSO candidates no matter what 
photometric properties are used.  Clearly optical spectra and/or emission 
line imaging are necessary to discriminate PNe from YSOs.

The sources we have classified as diffuse sometimes appear to be confined to 
relatively small areas when their IRAC colors are considered.  This is 
particularly true of their [5.8]$-$[8.0] color which appears to fall in an
narrow range of 1.5$<$[5.8]$-$[8.0]$<$2.2~mag.  As previously suggested in
\S~\ref{sec_diff} this is likely due to the dominance of PAH features and is 
consistent with our assessment that these are dusty knots of emission.  
Moreover, few of these sources have near-IR or 24~$\mu$m counterparts as
they do not appear in CMDs and CCDs that include longer or shorter wavelength
bands.  This is clearest in the bottom row of Figure~\ref{fig_CMDbyclass}
where nearly all the sources are marked with either an open circle 
or a cross because their natures are less certain.  This is perhaps a 
selection effect as these sources were in part classified as diffuse due 
to their previously mentioned lack of such near-IR and 24~$\mu$m counterparts.  
Higher angular resolution observations are needed for the diffuse sources to 
search for lower mass star formation hidden within these mid-IR knots. 

In the case of background galaxies, there sometimes appears to be a separation 
of the sources we have classified into two populations.  This is most clearly
evident in the [4.5]$-$[5.8] vs. [5.8]$-$[8.0] diagram in 
Figure~\ref{fig_CCDbyclass}.  One of these 
populations, those with [4.5]$-$[5.8]$<$[5.8]$-$[8.0]$-$1.0, appear to 
form a distinct group that can be excluded from being YSOs based upon their
mid-IR colors.  We have re-examined the SEDs of the sources classified as
background galaxies and find that the two populations follow the spectral 
classes found in \S~\ref{sec_bkggal} and shown in Figure~\ref{fig_exampleGAL}.
Both groups are visible in the SWIRE data supporting our
conclusion that both are extragalactic sources.  Furthermore, based on the 
results of \citet{Donleyetal08} the group that can not be easily separated 
from the YSO population is likely dominated by AGN and obscured AGN.

Overall, we find that such diagrams cannot be used to unambiguously determine 
the nature of an individual source.  It may yet be possible to determine a 
probability for the nature of a source from a combined analysis of all sources
in multiple CMDs and CCDs.  This cannot be achieved using the results of our 
current analysis because the only portion of color-magnitude space that has been 
completely analyzed is the wedge between our initial selection criteria in the 
[8.0] vs. [4.5]$-$[8.0] CMD.

\subsection{Comparison with Star Formation Results from the SAGE Team Analysis}

The SAGE collaboration has recently identified YSO candidates in the LMC 
using much of the same raw data we analyzed in this work \citep{Wetal08}.
The SAGE YSOs are initially identified using a complex set of color and 
magnitude criteria based on the predictions from a grid of radiative 
transfer models that attempt to simulate the emission from YSOs 
\citep{Wetal03a,Wetal03b,Retal06}.  This initial selection resulted in 3773 
candidates.  These candidates were then culled by requiring a YSO to: 
(1) have a modest amount of diffuse 24~$\mu$m emission at its location, 
(2) be detected in at least three bands among the four IRAC and 
MIPS~24$\mu$m bands, and (3) have an SED that could not be fit by a normal 
stellar atmosphere model.  This reduced the number of YSO candidates found 
by the SAGE team to 1197 which they further classified into YSOs, evolved 
stars, PNe and galaxies.

We have searched our photometric database for sources with positional matches
within 1\arcsec\ or the 1197 candidates found by the SAGE team.  
Table~\ref{tab_sagecomp1} summarizes the results of this comparison.  
Of the 1197 SAGE candidates, 1190 are included in our photometric catalog while 
7 had no counterpart.
Of the 1190 sources in our photometric catalog, 579 met the photometric 
criteria in the [8.0] vs. [4.5]$-$[8.0] CMD to be included in our initial 
YSO sample while the remaining 611 were excluded.
We have examined the photometric properties of these 611 sources 
to determine why they were excluded and found: 346 were excluded 
because they had [4.5]$-$[8.0]$<$2.0, 200 of the remaining sources were 
excluded because they had [8.0]$>$14$-$[4.5]$-$[8.0], and the remaining 
65 were excluded due to the absence of a flux measurement at either 
4.5~$\mu$m or 8.0~$\mu$m.  Therefore, 
$\sim$89\% of these 611 sources were excluded by the criteria used to 
prevent AGB, normal stars, and galaxies from heavily contaminating our 
sample while only $\sim$11\% did not have the necessary photometric 
measurements to be considered in our analysis.

We have further compared the classifications of the 579 sources in
common between our initial YSO sample selected from the 
[8.0] vs. [4.5]$-$[8.0] CMD with those assigned by the analysis of the
SAGE team.  The results are summarized in Table~\ref{tab_sagecomp2}.  
For cases where our classification differed we have reexamined our 
classification but found only a few cases where we had classified a 
source ``S" (a star with an IR excess) or ``SC" (a star with an IR excess 
or possibly a YSO) that might be reclassified as an evolved star or 
switched to a more probable YSO category.
The SAGE team had four classes which include the majority of their 
YSO candidate sources: YSO\_hp (high probability YSOs), 
YSOs, evolved stars, and PNe.  If we compare the sources classified as
YSO\_hp with our results we find that $\sim$76\% are present in one of
our three groups of YSOs but only $\sim$63\% are present among the
definite or probable YSOs.  Similarly, for the sources in the SAGE YSO
class, we find $\sim$69\% are present among our three YSO groups and 
only $\sim$61\% are present among our definite and probable YSOs.
For those sources classified as YSOs by the SAGE team, our analysis
has concluded that $\sim$4 to 12\% are likely to be evolved stars and
$\sim$20 to 33\% are likely background galaxies according to our criteria.
We find better agreement for sources \citet{Wetal08} classified as evolved 
stars and PNe, where we classify only $\sim$15\% of the evolved stars and 
$\sim$30\% of the PNe differently.

While the similarities are encouraging, the discrepancies between our results
and those of the SAGE analysis point to possible problems with both surveys.
In order to restrict the number of sources we had to classify, our analysis 
currently ignores many sources with [4.5]$-$[8.0]$<$2.0 and many very red 
sources with [8.0]$>$14$-$([4.5]$-$[8.0]).  The first cutoff may ignore 
many YSOs that are in the later stages of their evolution.  Similarly,
the second cutoff leaves a portion of color-magnitude space with [4.5]$-$[8.0]$>$3.5,
that does not appear to have a significant number of contaminating galaxies based on
the results from the SWIRE survey but which likely contains many fainter 
(and presumably less massive) YSOs. 
On the other hand, the \citet{Wetal08} sample appears to be contaminated by 
background galaxies by as much as 33\%.  In many of these cases we are certain of our 
diagnosis as the source was extended in multiple bands.  We suggest that the 
root of this problem is that the 
integrated SEDs of galaxies will generally show some star formation and thus
can often be well fit by SED models in the libraries constructed by
\citet{Retal06}.
The discrepancies among the sources classified as PNe by the SAGE analysis
may be due to the use of PN catalogs that include many sources that have not 
yet been spectroscopically confirmed \citep[e.g.,][]{RP06}.  Follow-up 
spectroscopic observations should be able to discern the true nature of these
sources.  The closest match we find between the two samples are among the 
evolved stars with $\sim$85\% concordance but there appears to be a fundamental
discrepancy among the sources we classified as EROs which \citet{Wetal08}
have always identified as YSOs.  

The largest difference between the SAGE catalog of YSOs in \citet{Wetal08}
and those presented 
in this paper is that we have identified 603 definite YSOs and 250 probable 
YSOs that do not appear in the SAGE YSO catalog.  Expressed differently, the 
SAGE YSO catalog misses $\sim$73\% of our YSOs.  This cannot be explained by
the color criteria used in the analysis by \citet{Wetal08} as it includes a 
broader portion of color--magnitude space.  Furthermore, if we look at
the brightness of the sources missed by \citet{Wetal08}, we find that the 
percentage of sources missed remains roughly constant (between $\sim$70 and 80\%)
over the brightness range of $5<[8.0]<12$ mag. \citet{Chen09} have analyzed 
the population of YSOs in the N\,44 region and find a similar percentage of 
sources missing from the \citet{Wetal08} catalog.  By comparing their source lists to 
the original SAGE photometry catalog, they conclude that many sources are 
excluded from the SAGE photometric catalog because of either
a higher signal--to--noise ratio threshold or a stricter requirement for a
point-source morphology.  The latter of these two possibilities is more consistent 
with our finding that the fraction of sources missed throughout the LMC does not 
vary with the brightness of the sources.

The photometric extraction method used for this work has allowed the inclusion of 
marginally extended sources and sources amid a complex interstellar 
background in our initial photometric catalogs.  This is extremely important for 
uncovering the YSO population in the LMC.  For Galactic searches, YSOs are generally 
resolved from their interstellar and even circumstellar surroundings.  In the LMC, 
where 1\arcsec$\simeq$0.25~pc, the circumstellar and interstellar surroundings 
will be much harder to separate from the sources of interest.  Furthermore, 
with the resolution afforded by {\it Spitzer} in many cases multiple YSOs are 
likely to be measured as a single source.  Thus, relaxing the search criteria 
for sources is not only justified but necessary to obtain 
the most complete census of YSOs possible.  In turn, this also requires that 
the individual sources be examined carefully in order to remove ``diffuse'' 
sources or local peaks in the dusty interstellar medium.

In neither the \citet{Wetal08} analysis nor the one we present here is the YSO
sample complete.  Our source lists should be relatively complete for the
region of the [8.0] vs. [4.5]$-$[8.0] CMD analyzed but will obviously miss
sources outside that region.  The \citet{Wetal08} analysis on the other hand 
will provide examples of some source classes we have not analyzed; however, 
their more automated system of analysis appears to miss many candidates in the region
of color-magnitude space we have exhaustively covered and includes sources
reject as background galaxies.  Studies that attempt to examine the spatial 
distribution of YSOs and compare 
with other tracers should be wary of the limitations of either YSO catalog.

\subsection{Luminosity Function of our YSO Candidates}

Figure~\ref{fig_YSOlumfunc} shows the mid-IR luminosity functions 
in observational units for the definite and probably YSO candidates from 
this paper.  For the IRAC bands the distribution of brighter sources appears 
to roughly follow a power law until the numbers of YSO candidates
begin to become incomplete due to the selection criteria that 
minimized the contamination by background galaxies.  In the MIPS~24~$\mu$m band 
the luminosity function appears much different.  This is because the numbers 
are likely incomplete over most of the plotted range where: (1) sources with 
[24]$\lesssim$1 will be saturated, (2) sources with [24]$\gtrsim$7.0 appear 
are incomplete because fainter sources are more difficult to identify 
amid diffuse emission, and (3) sources with [24]$\gtrsim$5.0 are likely 
incomplete due to the galaxy cutoff criteria.

We have assumed the rising portion of the YSO luminosity functions have a 
functional form $\log{N(m)}\propto\,a\,m$, where $m$ is the center of each 
magnitude bin, $N(m)$ is the number of sources in each bin.  Least-squares
fits to the luminosity functions find the slope, $a$, to range between
0.35 and 0.44 for the different IRAC bands with an average value of $\sim$0.4.
The individual results are summarized in columns 2 and 3 of 
Table~\ref{tab_lumfunc}.  In physical units the luminosity function has
the form $N \propto L^{b}$, where the power law index $b$ is related to the 
slope $a$ by $b=-2.5~a$.  Thus, the fitted slopes are roughly consistent with 
a luminosity function with power law of index $b=-1$.  If we assume a 
mass--luminosity relation, $L\propto M^{2.4}$, for massive stars 
\citep{Tetal96} and substitute this for $L$ in the luminosity function we 
find a mass function where $N \propto M^{-2.4}$.
While this result is remarkably consistent with those found for other
determinations of the initial mass function for stars with masses 
$\gtrsim10~M_\odot$ \citep{S55,MS79,Kroupa01}; it may also be a fortuitous 
coincidence as there are many potential problems in the analysis.  The 
most notable of these are: 
(1) the assumption that the YSO candidates are indeed single rather than
binary or multiple sources, 
(2) the assumption that the YSO candidates are indeed high- and intermediate 
mass YSOs, and 
(3) the unknown effect of the population of YSOs with $[4.5]-[8.0]<2.0$ that 
were not considered in our analysis.

The first two potential problems can be addressed by the results of a study 
of the YSO population in the LMC \ion{H}{2} complex N44 by \citet{Chen09}.  
This study supplemented {\it Spitzer} data with optical and near-IR 
observations that are more sensitive and have a higher spatial resolution 
than typically available for our sample.  They find that $\sim$60\% of the 
N44 YSO candidates show signs in the near-IR or optical of being either 
extended or composed of multiple sources but that most multiples had a 
single source that dominates the mid-IR emission.  Furthermore, they could 
establish whether or not the SEDs of their YSO candidates were contaminated
by emission from multiple sources prior to using the models of \citet{Retal06} 
to fit the SED.  Those results indeed suggest that sources with [8.0]$<$8.0~mag
are best fit with models where the central YSO has $\gtrsim$9~$M_\odot$ 
\citep[Table~7 of][]{Chen09} and suggest that our candidate YSOs are indeed 
progenitors of high- and intermediate-mass stars.  On the other hand, the 
fit results also demonstrate that models with a wide range of YSO masses 
are able to produce equally good fits to the SEDs.  Thus, a mass--luminosity 
relationship is not strictly valid when using a single IRAC band, however, it
may be a reasonable approximation when treating a large sample of YSOs.  

The analysis by the SAGE team affords us a means to explore the possible 
contribution of sources with $[4.5]-[8.0]<2.0$ to the mid-IR luminosity 
function.  \citet{Wetal08} have identified 268 YSOs with $[4.5]-[8.0]<2.0$ 
which should not be significantly contaminated by background galaxies over 
the brightness range where our mid-IR luminosity functions were complete.  
Therefore, using the flux measurements from Table~2 of \citet{Wetal08} we 
have supplemented our sample of YSOs to test whether there are significant 
changes to the fitted slopes ($a$).  Using the same brightness ranges as 
above we find only small changes for the fitted value of $a$ (columns 4 and 5 
of Table~\ref{tab_lumfunc}) and no change for the average value.  Therefore, 
if the SAGE YSOs with $[4.5]-[8.0]<2.0$ are at least a representative 
population then the exclusion of YSOs with $[4.5]-[8.0]<2.0$ should not 
have significantly affected the slopes of the YSO luminosity functions.

\section{Summary}

We have used {\it Spitzer} IRAC and MIPS observations to obtain mid-IR
photometric measurements of $>$3.5~million sources in the LMC.  Comparison
between the results from our photometry for the LMC and SWIRE survey 
with those of the SAGE and SWIRE teams, respectively, indicate that our 
measurements have roughly the same accuracy and completeness.  We have 
used the mid-IR photometry to search for high- and intermediate-mass 
YSOs in the LMC and have identified a sample of 2,910 sources for further 
consideration.

Subsequent analysis using images and photometry at optical, near-IR and 
mid-IR wavelengths have identified a sample of 1,172 definite and probable 
YSO candidates along with 213 other sources for which a YSO nature could 
not be definitively excluded.  In the process of identifying the YSOs we 
have also cataloged 117 objects which are likely obscured AGB stars and 
1075 objects that are most likely background galaxies.

Using our classification of sources we have shown that there are no simple 
diagnostics in color--magnitude space that can be used to uniquely separate 
this population of YSOs in the LMC from background galaxies, PNe, or AGB stars.  
Comparison with the analysis performed on the same dataset by the SAGE 
team \citep{Wetal08} has found that this previous analysis may be contaminated 
at a level of roughly 20--30\% by background galaxies and misses over 850 YSO 
candidates in the region where their analysis overlaps ours in color-magnitude 
space.  A simple analysis of the mid-IR luminosity function of the YSOs
suggest that the mass function of YSOs has a power-law index of $-$2.4, 
consistent with a Salpeter mass function, but it may simply be a fortuitous 
coincidence that this value is similar to determinations of the stellar IMF 
for massive stars in the Solar Neighborhood.  

Clearly the nature and masses of these candidate YSOs must be better 
established before they can be used to rigorously study the star formation 
process or establish an independent estimate of the YSO mass-function in 
the LMC.  To this end, {\it Spitzer} program 40650 has recently obtained 
follow-up mid-IR spectroscopic observations of 269 of the brighter definite 
and probable YSO presented in this paper and confirm that $\gtrsim$95\% of
these sources are indeed YSOs \citep{Setal09}.  Further follow-up observations with
{\it Herschel} should allow better estimates of the source masses and 
evolutionary states.  Deep, high-resolution, near- 
and/or mid-IR observations will enable an accurate assessment of the 
multiplicity of these YSO candidates and enable a search for associated 
lower-mass star formation.

\acknowledgments 
This research was supported by NASA grants JPL1264494 and JPL1290956 and
through the NSF grant AST08-07323.  We thank the anonymous referee for their
suggestions which have helped improve this paper.  We also thank S.~D.~Points and 
A.~Rest for obtaining some of the ISPI and MOSAIC observations used in this work.  
This publication has made use of the data products from the Two Micron All 
Sky Survey, which is a joint project of the University of Massachusetts and the 
Infrared Processing and Analysis Center/California Institute of Technology, 
funded by NASA and the NSF.  The Digitized Sky Surveys images used were 
produced at the Space Telescope Science Institute under U.S. Government 
grant NAG W-2166. The images of these surveys are based on photographic 
data obtained using the UK Schmidt Telescope. The plates were processed 
into the present compressed digital form with the permission of these 
institutions.  Furthermore, this research has made use of the SIMBAD 
database, operated at CDS, Strasbourg, France and SAOImage DS9, developed 
by Smithsonian Astrophysical Observatory.

\clearpage

\begin{figure}
\epsscale{0.9}
\plotone{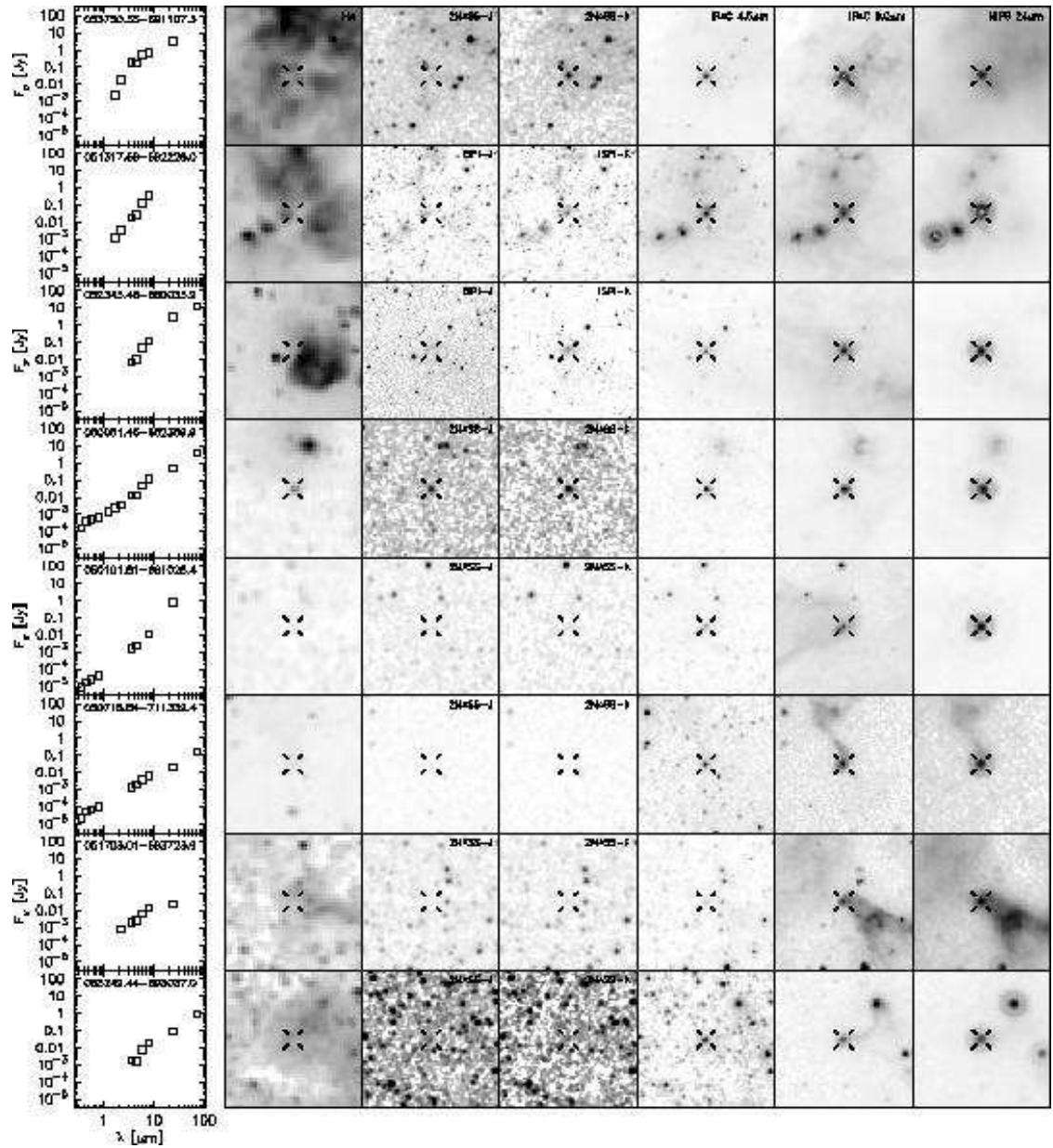}
\figurenum{13a}
\caption{A few examples of YSO candidates that have been classified as 
``Definite.''  The panels in the left column show the SEDs for each source.
The images in each row are MCELS-H$\alpha$+continuum, 2MASS-J (or ISPI-J if available), 
2MASS-K (or ISPI-K if available), IRAC 4.5~$\mu$m, IRAC 8.0~$\mu$m, and MIPS 
24~$\mu$m (from left to right) and show a 2$^\prime$ field of view centered 
on the source (marked).
Figures~13b-–x present SEDs and images for all 233 sources 
classified as definite YSOs with [8.0]$<$8 and are available in the
electronic edition of the Journal.}
\label{fig_exampleYSO1}
\end{figure}

\clearpage

\begin{figure}
\epsscale{0.9}
\plotone{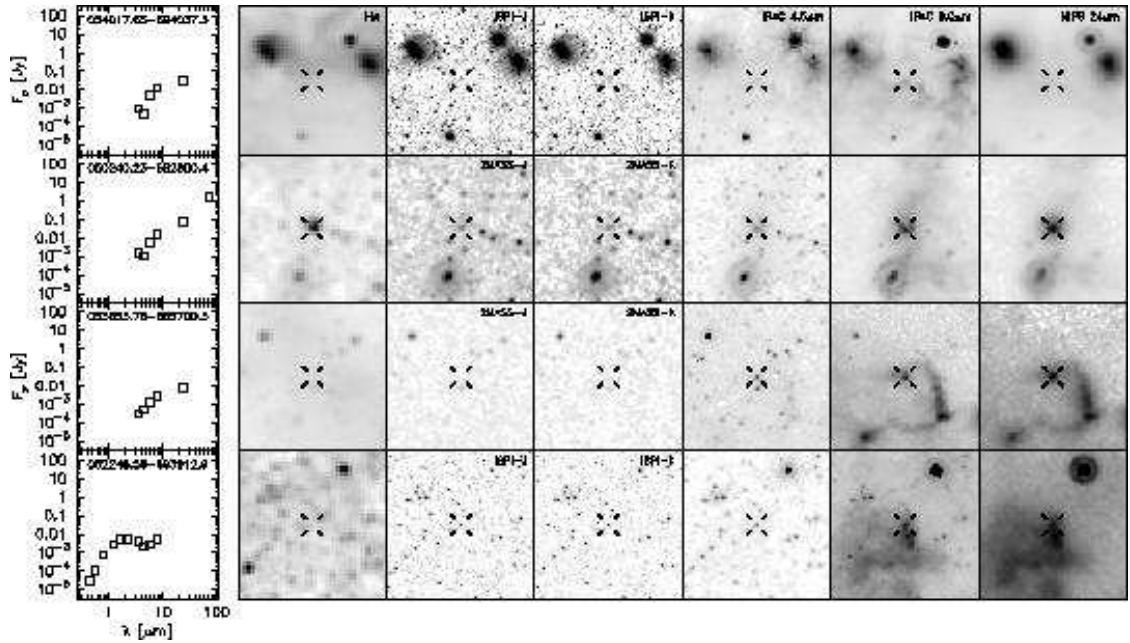}
\figurenum{14a}
\caption{Similar to Figure~\ref{fig_exampleYSO1} but for YSO candidates that 
have been classified as ``Probable.''  From top to bottom the rows show the
SEDs and images of 054017.65$-$694037.3, 050240.23$-$682800.4, 
053653.78$-$665700.3, and 052248.56$-$693912.9, which are respectively examples 
of sources in the CD, CG, CG, and CS classes.  Figures~\ref{fig_exampleYSO2}b\&c
present SEDs and images for the 14 sources classified as probable YSOs with
[8.0]$<$8 and are available in the electronic edition of the Journal.}
\label{fig_exampleYSO2}
\end{figure}



\setcounter{figure}{14}
\begin{figure}
\epsscale{0.9}
\plotone{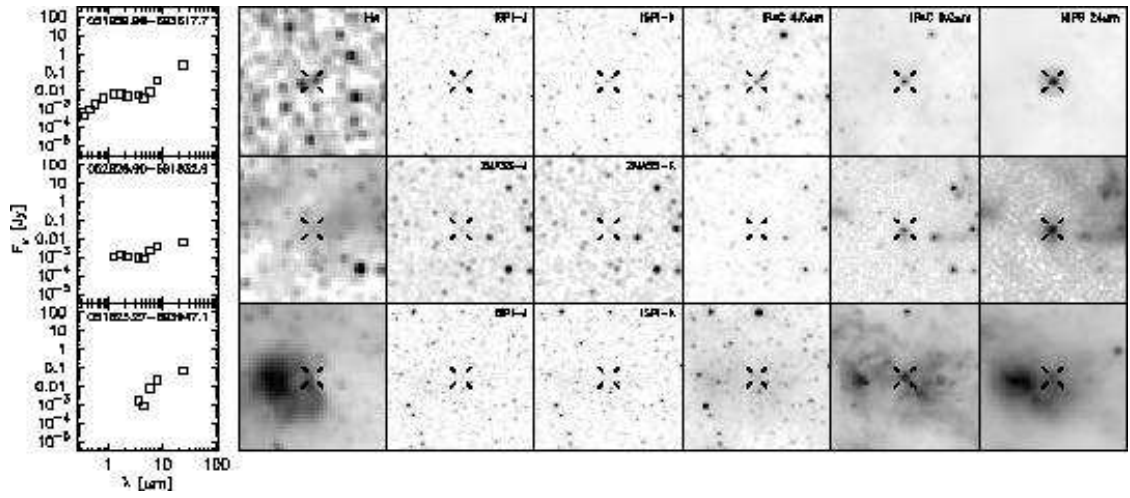}
\caption{Similar to Figure~\ref{fig_exampleYSO1} but for candidates that 
have been classified as ``Possible.'' From top to bottom the rows show the
SEDs and images of 051959.98$-$693617.7, 052825.90$-$691832.9, 
051823.27$-$693947.1, which are respectively examples of the SC, GC, and 
DC classes.}
\label{fig_exampleYSO3}
\end{figure}

\begin{figure}
\plotone{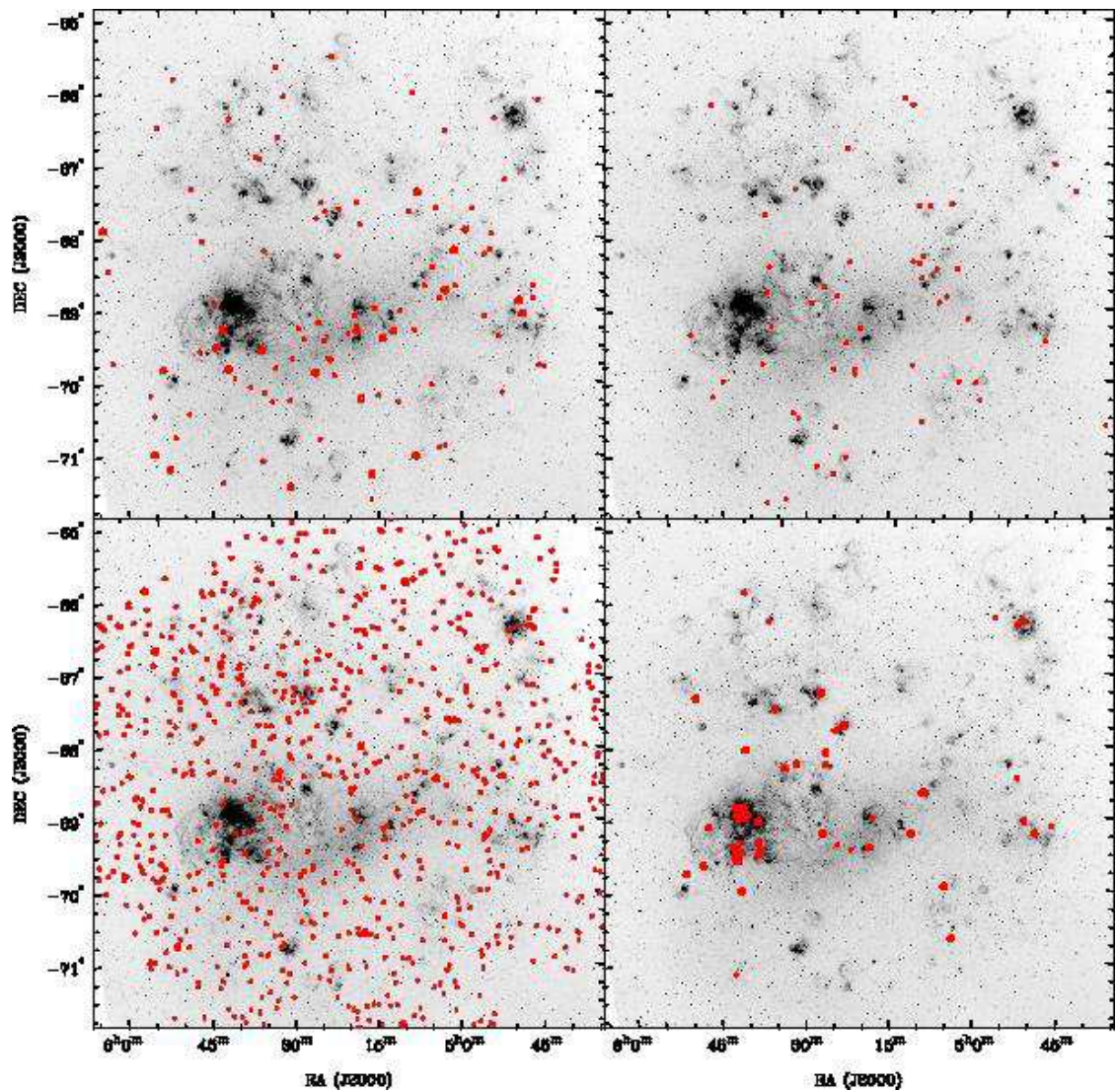}
\caption{Each panels shows the MCELS H$\alpha$+continuum image of the LMC overlaid 
with the positions of sources that have been excluded from our sample of 
candidate YSOs.  The positions of evolved stars ({\it top left}), 
planetary nebulae ({\it top right}), background galaxies ({\it bottom left}), 
and diffuse sources ({\it bottom right}) are marked with red filled circles when 
the nature of the source is relatively more certain and red open circles if their
classification should be considered questionable.} 
\label{fig_distOTHER}
\notetoeditor{This figure should appear in color in the print and electronic edition of the Journal.}
\end{figure}

\begin{figure}
\plotone{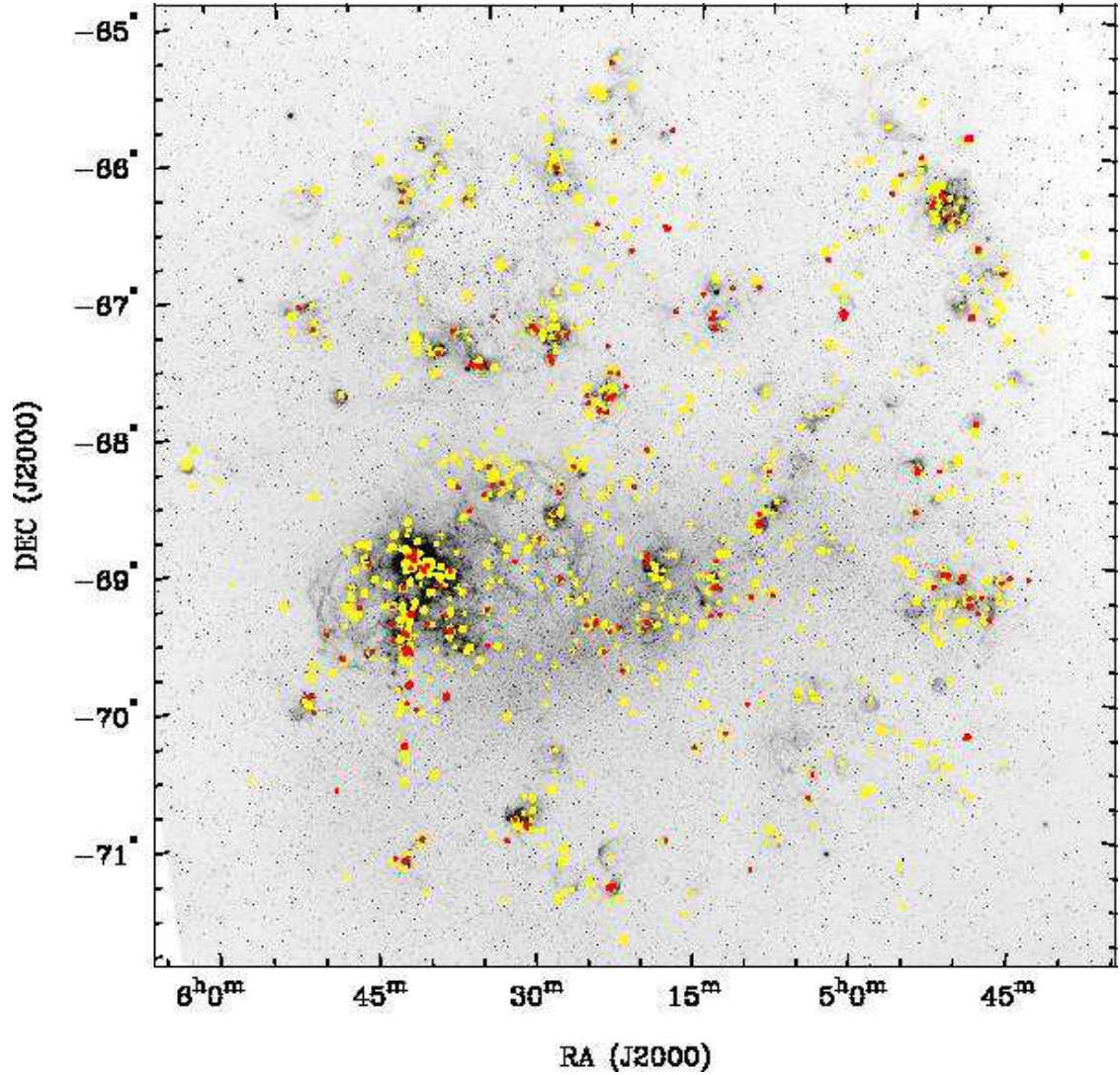}
\caption{MCELS H$\alpha$+continuum image of the LMC overlaid with positions of 
candidate YSOs.  Sources shown with red symbols have [8.0]$<$8.0~mag 
roughly indicating more massive YSO candidates while the remaining 
YSO candidates are shown with yellow symbols.  Solid circles mark the 
locations of source classified as ``Definite YSOs,'' open circles mark 
``Probable YSOs,'' and crosses mark ``Possible YSOs.''}
\label{fig_distYSO}
\notetoeditor{This figure should appear in color in the print and electronic edition of the Journal.}
\end{figure}

\begin{figure}
\plotone{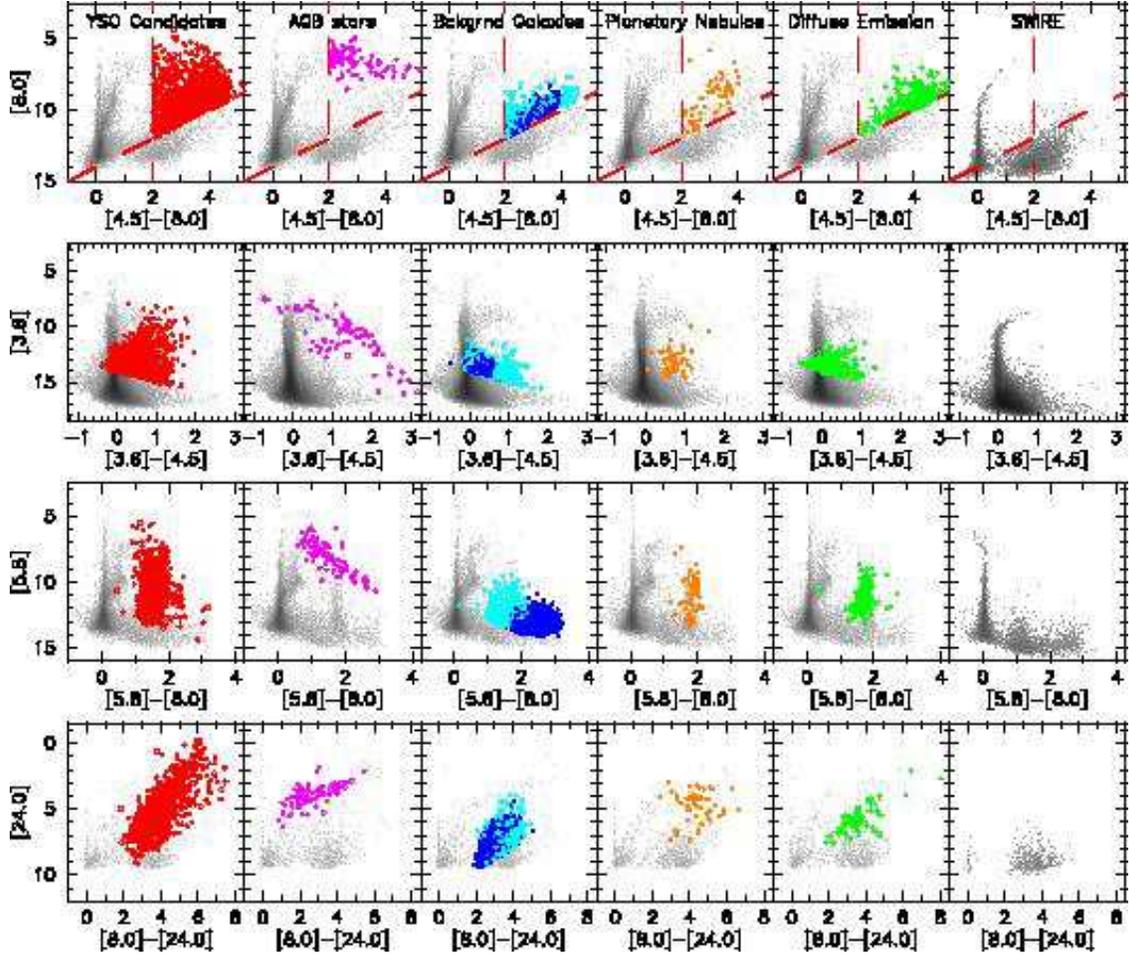}
\caption{Each row shows a different CMD where the first five columns show
a Hess diagram (greyscale) using all sources in the LMC.  Each column 
is used for an individual source class indicated in the top row.
Solid circles are used when the nature of a source was considered highly
certain, open circles are used when the nature of a source is considered
probable, and crosses are used to indicate points that may have a more likely
explanation as to their nature.  The symbols also have a color based on their
class: YSOs (red), AGB stars (magenta), background galaxies (blue or cyan), 
PNe (orange), diffuse sources (green).  For the background galaxies, the blue symbols 
indicate a color of [4.5]$-$[5.8]$<$[5.8]$-$[8.0]$-$1.0, and the cyan symbols show 
the opposite.   The rightmost column in each row shows a Hess diagram derived from 
data in the SWIRE survey which should be dominated by Galactic stars and distant 
galaxies.  The dashed lines in the top row show the criteria used to select the 
initial sample of sources.}
\label{fig_CMDbyclass}
\notetoeditor{This figure should appear in color in the print and electronic edition of the Journal.}
\end{figure}

\begin{figure}
\plotone{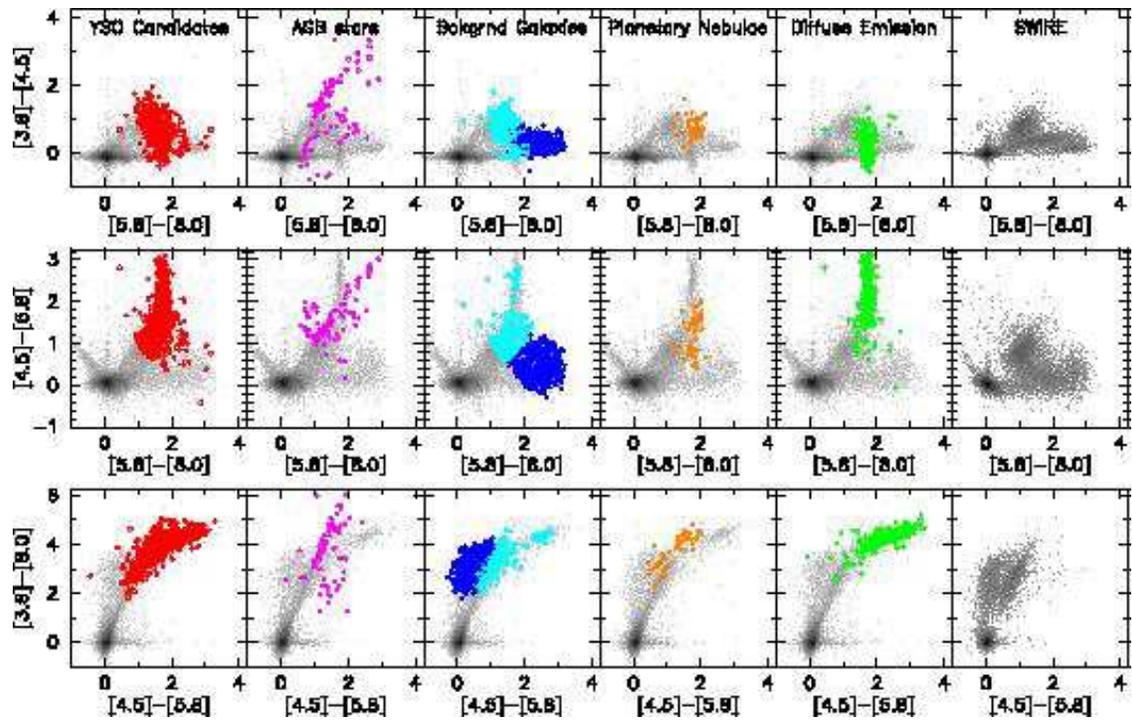}
\caption{Same as Figure~\ref{fig_CMDbyclass} but showing color-color diagrams.}
\label{fig_CCDbyclass}
\notetoeditor{This figure should appear in color in the print and electronic edition of the Journal.}
\end{figure}

\begin{figure}
\plotone{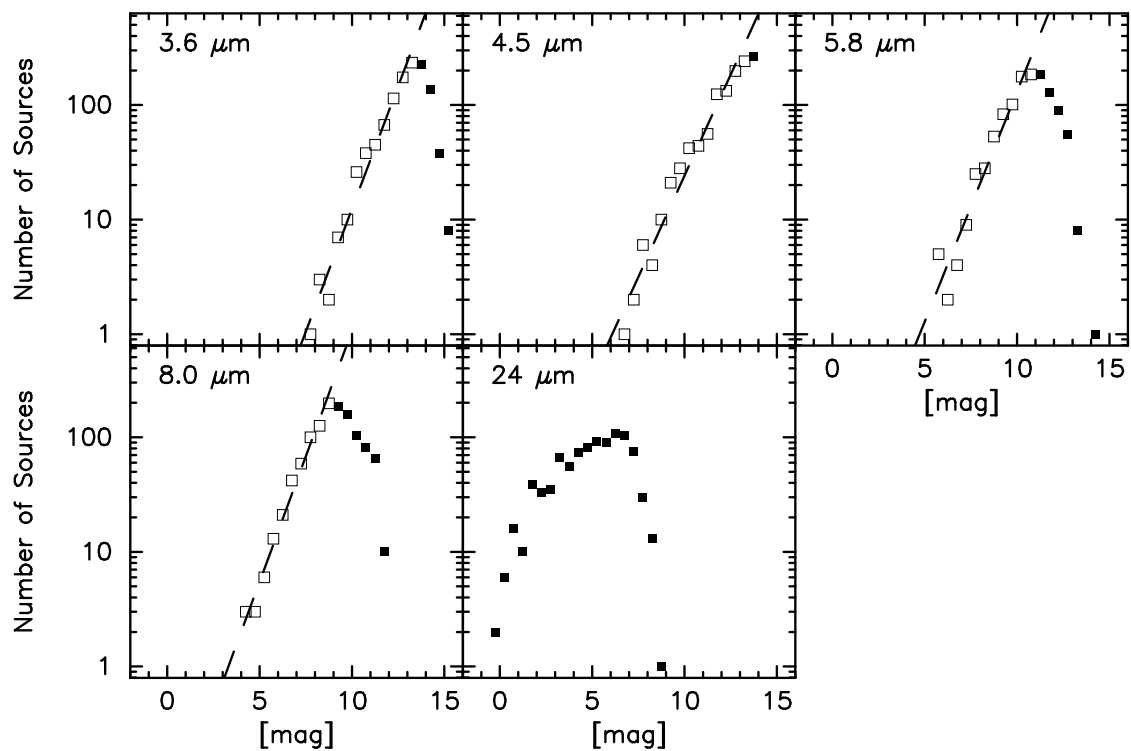}
\caption{Luminosity functions for the IRAC and MIPS~24~$\mu$m band constructed 
using the sources identified as definite or probable YSOs.  Open symbols indicate
measurements that were used to fit the slope for the brighter sources while filled 
symbols indicate measurements that were excluded when fitting.  The best fit for
each of the IRAC bands has been plotted as a dashed line.}
\label{fig_YSOlumfunc}
\end{figure}

\clearpage

\begin{deluxetable}{rllrc}
\tabletypesize{\footnotesize}
\tablecolumns{5}
\tablewidth{0pt}
\tablecaption{Observation Summary\label{observe_tab}}
\tablehead{
\colhead{Program} & Principle    & \colhead{}  & \multicolumn{2}{c}{Typical Observation Parameters} \\
\colhead{ID}      & Investigator & Region(s)   & \colhead{IRAC} & \colhead{MIPS} \\
\hline
\multicolumn{5}{c}{Observations Used for Photometry and Imaging}
}
\startdata  
 124  & Gehrz      & N\,159, N\,160                     & 10$\times$12\tablenotemark{1}s &       \\
1032  & Brandl     & 30\,Dor                            &  3$\times$12\tablenotemark{1}s &       \\
1061  & Gorjian    & N\,206                             &  3$\times$12\tablenotemark{1}s &       \\
3565  & Chu        & N\,11, N\,44, N\,51, N\,63         &  5$\times$30\tablenotemark{1}s & Med Scan \\
      &            & N\,70, N\,144, N\,180              &                                &       \\
20203 & Meixner    & SAGE Legacy Program                &  4$\times$12\tablenotemark{1}s & 2$\times$Fast Scan \\
\cutinhead{Observations Used for Imaging Only} 
  34  & Fazio      & AB\,Dor                            &  5$\times$30\tablenotemark{1}s &       \\
  63  & Houck      & 30\,Dor                            &  3$\times$12\tablenotemark{1}s &       \\
 118  & Werner     & 5 MACHO Events                     & 24$\times$30\tablenotemark{\phn}s &    \\
 124  & Gehrz      & SN\,1987A                          & 64$\times$12\tablenotemark{\phn}s &    \\
      &            & N\,49                              & 50$\times$2\tablenotemark{\phn}s  &    \\
 125  & Fazio      & LMC\,Bar                           &  4$\times$12\tablenotemark{1}s &       \\ 
 249  & Indebetouw & LMC-C and LMC-S fields             &                & Fast Scan \\
 717  & Rieke      & N\,206                             &                & Med Scan \\
3400  & Rich       & NGC\,1786, NGC\,1806, NGC\,1831    & 72$\times$12\tablenotemark{1}s &       \\
      &            & NGC\,1856, NGC\,1866, NGC\,2004    &                &          \\
      &            & NGC\,2173, NGC\,2210, NGC\,2004    &                &          \\
3483  & Rho        & N\,132D                            &  9$\times$30\tablenotemark{\phn}s &    \\
3578  & Misselt    & 32 LMC Extinction Probes           &  3$\times$12\tablenotemark{\phn}s &    \\
3680  & Borkowski  & 15 LMC\,SNRs                       &  5$\times$30\tablenotemark{\phn}s &    \\
3725  & Goudfrooij & NGC\,1751, NGC\,1783, NGC\,1846    & 14$\times$100\tablenotemark{\phn}s &   \\
      &            & NGC\,1978, NGC\,1987, NGC\,2108    &                &          \\
      &            & NGC\,2808                          &                &          
\enddata
\tablenotetext{1}{Observation used high dynamic range mode and have complementary short exposures.}
\end{deluxetable}

\begin{deluxetable}{cccccc}
\tablewidth{0pt}
\tablecaption{Photometric Extraction Parameters\label{phot_tab}}
\tablehead{
\colhead{}           & \colhead{Aperture}  & \colhead{Sky}       & \colhead{Aperture}   & \colhead{Zero}  & Assumed Flux \\
\colhead{Wavelength} & \colhead{Radius}    & \colhead{Annulus}   & \colhead{Correction} & \colhead{Point} & Calibration  \\
\colhead{[$\mu$m]}   & \colhead{[\arcsec]} & \colhead{[\arcsec]} & \colhead{}           & \colhead{[Jy]}  & Accuracy     
}
\startdata
  3.6 &  3.6 & 3.6--8.4   & 1.124 & 277.5  & 5\%  \\ 
  4.5 &  3.6 & 3.6--8.4   & 1.127 & 197.5  & 5\%  \\ 
  5.8 &  3.6 & 3.6--8.4   & 1.143 & 116.6  & 5\%  \\ 
  8.0 &  3.6 & 3.6--8.4   & 1.234 &  63.1  & 5\%  \\ 
 24.0 &  6.0 & 15.0--23.0 & 1.798 &  7.14  & 10\%  \\ 
 70.0 & 18.0 & 18.0--39.0 & 1.927 &  0.775 & 20\%  
\enddata
\end{deluxetable}

\begin{deluxetable}{rrrrrrrrrrrrrrrr}
\tablecolumns{16}
\rotate
\tablewidth{0pt}
\tabletypesize{\scriptsize}
\tablecaption{Detailed Comparison of Our Photometric Measurements with SAGE and SWIRE Results\label{compphot_tab}}
\tablehead{    
\colhead{ }  & 
\multicolumn{3}{c}{IRAC 3.6~$\mu$m} &
\multicolumn{3}{c}{IRAC 4.5~$\mu$m} &
\multicolumn{3}{c}{IRAC 5.8~$\mu$m} &
\multicolumn{3}{c}{IRAC 8.0~$\mu$m} &
\multicolumn{3}{c}{MIPS 24.0~$\mu$m} \\
\colhead{Range} &
\colhead{N$_{3.6}$}  & \colhead{$\Delta_{3.6}$\tablenotemark{a}} & \colhead{$\sigma_{3.6}$} &
\colhead{N$_{4.5}$}  & \colhead{$\Delta_{4.5}$\tablenotemark{a}} & \colhead{$\sigma_{4.5}$} &
\colhead{N$_{5.8}$}  & \colhead{$\Delta_{5.8}$\tablenotemark{a}} & \colhead{$\sigma_{5.8}$} &
\colhead{N$_{8.0}$}  & \colhead{$\Delta_{8.0}$\tablenotemark{a}} & \colhead{$\sigma_{8.0}$} &
\colhead{N$_{24.0}$} & \colhead{$\Delta_{24.0}$\tablenotemark{a}} & \colhead{$\sigma_{24.0}$} \\
\colhead{[mag]} & 
\colhead{}  & \colhead{[mag]} & \colhead{[mag]} &
\colhead{}  & \colhead{[mag]} & \colhead{[mag]} &
\colhead{}  & \colhead{[mag]} & \colhead{[mag]} &
\colhead{}  & \colhead{[mag]} & \colhead{[mag]} &
\colhead{}  & \colhead{[mag]} & \colhead{[mag]} \\
\hline
\multicolumn{16}{c}{Comparison with SWIRE results} 
}
\startdata
   5.0--6.0 &         &        &       &        &        &       &        &        &       &       &        &       &    12 & -0.016 & 0.013  \\
   6.0--6.0 &         &        &       &        &        &       &     10 & -0.109 & 0.026 &     7 & -0.171 & 0.062 &    29 & -0.014 & 0.030  \\
   7.0--8.0 &         &        &       &      9 & -0.087 & 0.033 &     18 & -0.045 & 0.030 &    21 & -0.048 & 0.061 &   107 & -0.019 & 0.034  \\
   8.0--9.0 &      27 & -0.157 & 0.087 &     29 & -0.046 & 0.044 &     32 & -0.025 & 0.031 &    33 & -0.036 & 0.030 &   591 & -0.026 & 0.054  \\
  9.0--10.0 &      66 & -0.052 & 0.078 &     73 & -0.021 & 0.058 &     70 & -0.021 & 0.040 &    84 & -0.032 & 0.047 &   937 & -0.009 & 0.074  \\
 10.0--11.0 &     165 & -0.017 & 0.052 &    152 & -0.003 & 0.041 &    155 & -0.020 & 0.040 &   227 & -0.039 & 0.103 & \\
 11.0--12.0 &     329 & -0.006 & 0.053 &    336 & -0.009 & 0.047 &    359 & -0.019 & 0.064 &   663 & -0.052 & 0.096 & \\
 12.0--13.0 &     718 & -0.011 & 0.063 &    718 & -0.014 & 0.068 &    778 & -0.026 & 0.106 &  1903 & -0.062 & 0.073 & \\
 13.0--14.0 &    1389 & -0.016 & 0.073 &   1496 & -0.019 & 0.076 &   1617 & -0.029 & 0.129 &  4142 & -0.057 & 0.107 & \\
 14.0--15.0 &    3425 & -0.029 & 0.073 &   4343 & -0.036 & 0.080 &   4198 & -0.027 & 0.195 &  1574 &  0.004 & 0.209 & \\
 15.0--16.0 &   11583 & -0.041 & 0.087 &  15491 & -0.037 & 0.105 &   3120 &  0.132 & 0.265 &     6 &  1.023 & 0.731 & \\
 16.0--17.0 &   37568 & -0.029 & 0.118 &  46532 & -0.016 & 0.152 &     98 &  0.766 & 0.249 &       &        &       & \\
 17.0--18.0 &   45100 & -0.006 & 0.164 &  17339 &  0.054 & 0.220 &        &        &       &       &        &       & \\
\cutinhead{Comparison with SAGE results} 
   1.0--2.0 &         &        &       &        &        &       &        &        &       &       &        &       &    44 & -0.046 & 0.226  \\
   2.0--3.0 &         &        &       &        &        &       &        &        &       &       &        &       &   111 & -0.004 & 0.221  \\
   3.0--4.0 &         &        &       &        &        &       &     10 & -0.007 & 0.047 &    15 &  0.000 & 0.046 &   264 &  0.005 & 0.315  \\
   4.0--5.0 &         &        &       &        &        &       &     23 &  0.003 & 0.020 &    37 &  0.007 & 0.073 &   457 & -0.007 & 0.299  \\
   5.0--6.0 &         &        &       &     36 &  0.001 & 0.059 &     81 &  0.004 & 0.124 &   155 &  0.010 & 0.148 &   838 & -0.024 & 0.327  \\
   6.0--7.0 &     163 &  0.019 & 0.078 &    134 & -0.002 & 0.165 &    269 &  0.022 & 0.187 &   607 &  0.010 & 0.132 &  1586 & -0.061 & 0.375  \\
   7.0--8.0 &     440 &  0.021 & 0.209 &    655 &  0.000 & 0.284 &   1063 &  0.023 & 0.216 &  1168 &  0.013 & 0.189 &  3468 & -0.094 & 0.366  \\
   8.0--9.0 &    1764 &  0.013 & 0.248 &   1958 &  0.003 & 0.205 &   1927 &  0.013 & 0.166 &  2533 &  0.024 & 0.095 & 10871 & -0.091 & 0.262  \\
  9.0--10.0 &    6553 &  0.009 & 0.160 &   5347 &  0.010 & 0.163 &   5768 &  0.000 & 0.088 &  7110 &  0.010 & 0.082 & 14176 & -0.103 & 0.176  \\
 10.0--11.0 &   13430 &  0.008 & 0.140 &  12371 &  0.010 & 0.082 &  13847 &  0.003 & 0.102 & 13470 &  0.011 & 0.102 & \\
 11.0--12.0 &   27605 & -0.002 & 0.128 &  24797 &  0.001 & 0.110 &  25765 &  0.008 & 0.133 & 27413 &  0.013 & 0.133 & \\
 12.0--13.0 &   91654 & -0.006 & 0.120 &  84025 & -0.001 & 0.112 &  87999 &  0.008 & 0.152 & 90055 &  0.015 & 0.131 & \\
 13.0--14.0 &  179618 & -0.015 & 0.158 & 174575 & -0.005 & 0.157 & 163275 &  0.010 & 0.148 & 62386 &  0.069 & 0.147 & \\
 14.0--15.0 &  352594 & -0.037 & 0.220 & 335183 & -0.021 & 0.197 & 107277 &  0.081 & 0.179 &       &        &       & \\
 15.0--16.0 &  691393 & -0.074 & 0.251 & 566721 & -0.023 & 0.190 &    371 &  0.469 & 0.203 &       &        &       & \\
 16.0--17.0 & 1059677 & -0.018 & 0.200 & 248425 &  0.070 & 0.174 &        &        &       &       &        &       & 
\enddata
\tablenotetext{a}{Differences are given as the value in the SAGE or SWIRE catalog minus the value determined in this work.}
\end{deluxetable}

\begin{deluxetable}{lrrrrrrrrrcl}
\rotate
\tablewidth{0pt}
\tabletypesize{\scriptsize}
\tablecaption{LMC Evolved Stars\label{tab_evolvedstar}}
\tablehead{
\colhead{Source ID} & 
\colhead{m$_{J}$} & \colhead{m$_{H}$} & \colhead{m$_{K_{\rm s}}$} &
\colhead{m$_{3.6}$} & \colhead{m$_{4.5}$} & \colhead{m$_{5.8}$} & 
\colhead{m$_{8.0}$} & \colhead{m$_{24.0}$} & \colhead{m$_{70.0}$} &
\colhead{}      & \colhead{Cross} \\
\colhead{ } & 
\colhead{[mag]} & \colhead{[mag]} & \colhead{[mag]} &
\colhead{[mag]} & \colhead{[mag]} & \colhead{[mag]} & 
\colhead{[mag]} & \colhead{[mag]} & \colhead{[mag]} &
\colhead{Class} & \colhead{Identification} 
}
\startdata 
 043257.40$-$692633.5 &    \nodata     &    \nodata     & 14.32$\pm$.07  & 10.43$\pm$.06  &  8.96$\pm$.05  &  7.72$\pm$.06  &  6.60$\pm$.05  &  4.67$\pm$.12 &  $>$4.19      & A &   MSX\,LMC\,1008 \\
 044827.63$-$695323.0 &    \nodata     &    \nodata     &    \nodata     & 10.85$\pm$.05  &  9.09$\pm$.06  &  7.84$\pm$.06  &  6.63$\pm$.05  &  4.43$\pm$.12 &   \nodata     & A &   MSX\,LMC\,1137 \\
 044858.14$-$684742.2 & 16.11$\pm$.09  & 15.04$\pm$.08  & 14.01$\pm$.07  & 11.55$\pm$.05  & 10.42$\pm$.05  &  9.34$\pm$.05  &  8.21$\pm$.06  &  5.52$\pm$.13 &  $>$4.09      & AS &  \\
 044918.53$-$695314.5 & 12.66$\pm$.03  & 10.85$\pm$.03  &  9.43$\pm$.02  &  7.60$\pm$.05  &  8.22$\pm$.05  &  7.07$\pm$.05  &  5.62$\pm$.05  &  4.05$\pm$.12 &   \nodata     & A &   MSX\,LMC\,1130 \\
 045128.99$-$685750.1 &  9.93$\pm$.02  &  8.62$\pm$.02  &  7.91$\pm$.01  &  7.59$\pm$.05  &  8.24$\pm$.05  &  6.78$\pm$.05  &  5.69$\pm$.05  &  3.25$\pm$.11 &  $>$2.84      & A &   MSX\,LMC\,1190 \\

\enddata
\tablecomments{Table~\ref{tab_evolvedstar} is presented in its entirety in the
electronic edition of the Astrophysical Journal.}
\tablecomments{Sources marked ``sat.'' at 24~$\mu$m are saturated or amid saturated diffuse emission.}
\end{deluxetable}

\begin{deluxetable}{lrrrrrrrrrcl}
\rotate
\tablewidth{0pt}
\tabletypesize{\scriptsize}
\tablecaption{LMC Planetary Nebulae\label{tab_pne}}
\tablehead{
\colhead{Source ID} & 
\colhead{m$_{J}$} & \colhead{m$_{H}$} & \colhead{m$_{K_{\rm s}}$} &
\colhead{m$_{3.6}$} & \colhead{m$_{4.5}$} & \colhead{m$_{5.8}$} & 
\colhead{m$_{8.0}$} & \colhead{m$_{24.0}$} & \colhead{m$_{70.0}$} &
\colhead{}      & \colhead{Cross} \\
\colhead{ } & 
\colhead{[mag]} & \colhead{[mag]} & \colhead{[mag]} &
\colhead{[mag]} & \colhead{[mag]} & \colhead{[mag]} & 
\colhead{[mag]} & \colhead{[mag]} & \colhead{[mag]} &
\colhead{Class} & \colhead{Identification} 
}
\startdata
 043834.77$-$703643.3 & 15.43$\pm$.06  &    \nodata     &    \nodata     & 12.35$\pm$.05  & 11.52$\pm$.05  & 10.11$\pm$.05  &  8.24$\pm$.06  &  3.99$\pm$.12 &  3.13$\pm$.24 & P &   SMP\,LMC\,1 \\
 044056.68$-$674802.4 &    \nodata     &    \nodata     &    \nodata     & 13.58$\pm$.06  & 12.78$\pm$.06  & 12.02$\pm$.06  & 10.37$\pm$.06  &  7.36$\pm$.17 &   \nodata     & P &   SMP\,LMC\,2 \\
 044808.56$-$672606.9 & 16.49$\pm$.13  & 15.87$\pm$.20  & 15.19$\pm$.21  & 13.15$\pm$.06  & 12.24$\pm$.05  & 11.51$\pm$.06  &  9.88$\pm$.05  &  6.65$\pm$.15 &  $>$4.53      & P &   SMP\,LMC\,5 \\
 045013.15$-$693356.9 & 15.68$\pm$.08  & 15.78$\pm$.16  & 14.35$\pm$.10  & 12.18$\pm$.05  & 10.98$\pm$.05  &  9.52$\pm$.05  &  7.76$\pm$.05  &  3.27$\pm$.11 &   \nodata     & P &   SMP\,LMC\,8 \\
 045137.83$-$670517.2 &    \nodata     &    \nodata     & 14.41$\pm$.09  & 10.39$\pm$.06  &  8.80$\pm$.05  &  7.30$\pm$.06  &  5.80$\pm$.05  &  2.94$\pm$.11 &  2.26$\pm$.22 & P &   SMP\,LMC\,11 \\

\enddata
\tablecomments{Table~\ref{tab_pne} is presented in its entirety in the
electronic edition of the Astrophysical Journal.}
\end{deluxetable}

\begin{deluxetable}{lrrrrrrrrrcl}
\rotate
\tablewidth{0pt}
\tabletypesize{\scriptsize}
\tablecaption{Background Galaxies\label{tab_gal}}
\tablehead{
\colhead{Source ID} & 
\colhead{m$_{J}$} & \colhead{m$_{H}$} & \colhead{m$_{K_{\rm s}}$} &
\colhead{m$_{3.6}$} & \colhead{m$_{4.5}$} & \colhead{m$_{5.8}$} & 
\colhead{m$_{8.0}$} & \colhead{m$_{24.0}$} & \colhead{m$_{70.0}$} &
\colhead{}      & \colhead{Cross} \\
\colhead{ } & 
\colhead{[mag]} & \colhead{[mag]} & \colhead{[mag]} &
\colhead{[mag]} & \colhead{[mag]} & \colhead{[mag]} & 
\colhead{[mag]} & \colhead{[mag]} & \colhead{[mag]} &
\colhead{Class} & \colhead{Identification} 
}
\startdata
 042257.86$-$693626.6 &    \nodata     &    \nodata     &    \nodata     & 14.85$\pm$.07  & 13.84$\pm$.06  &    \nodata     & 11.77$\pm$.06  &    \nodata    &   \nodata     & G &  \\
 042314.28$-$693012.4 & 16.19$\pm$.17  & 15.56$\pm$.20  & 14.62$\pm$.12  & 13.31$\pm$.06  & 12.77$\pm$.06  & 11.99$\pm$.06  & 10.08$\pm$.06  &    \nodata    &   \nodata     & G &  \\
 042315.11$-$693014.1 & 16.35$\pm$.14  & 15.82$\pm$.24  & 14.53$\pm$.10  & 13.58$\pm$.06  & 13.33$\pm$.06  & 12.56$\pm$.08  &  9.89$\pm$.06  &    \nodata    &   \nodata     & G &  \\
 042410.07$-$691308.9 &    \nodata     &    \nodata     &    \nodata     & 14.30$\pm$.06  & 13.84$\pm$.07  & 13.17$\pm$.09  & 11.45$\pm$.06  &  8.71$\pm$.23 &   \nodata     & G &  \\
 042451.48$-$685617.1 &    \nodata     &    \nodata     &    \nodata     & 14.39$\pm$.06  & 13.54$\pm$.06  & 12.55$\pm$.07  & 11.34$\pm$.07  &  7.61$\pm$.18 &   \nodata     & G &  \\

\enddata
\tablecomments{Table~\ref{tab_gal} is presented in its entirety in the
electronic edition of the Astrophysical Journal.}
\end{deluxetable}

\begin{deluxetable}{lrrrrrrrrrcl}
\rotate
\tablewidth{0pt}
\tabletypesize{\scriptsize}
\tablecaption{Probable Background Galaxies\label{tab_pgal}}
\tablehead{
\colhead{Source ID} & 
\colhead{m$_{J}$} & \colhead{m$_{H}$} & \colhead{m$_{K_{\rm s}}$} &
\colhead{m$_{3.6}$} & \colhead{m$_{4.5}$} & \colhead{m$_{5.8}$} & 
\colhead{m$_{8.0}$} & \colhead{m$_{24.0}$} & \colhead{m$_{70.0}$} &
\colhead{}      & \colhead{Cross} \\
\colhead{ } & 
\colhead{[mag]} & \colhead{[mag]} & \colhead{[mag]} &
\colhead{[mag]} & \colhead{[mag]} & \colhead{[mag]} & 
\colhead{[mag]} & \colhead{[mag]} & \colhead{[mag]} &
\colhead{Class} & \colhead{Identification} 
}
\startdata
 044808.84$-$684219.4 & 16.96$\pm$.20  & 15.64$\pm$.15  &    \nodata     & 13.74$\pm$.06  & 12.78$\pm$.06  & 11.83$\pm$.06  & 10.19$\pm$.06  &  6.47$\pm$.15 &  1.73$\pm$.22 & GC &  \\
 044853.98$-$695856.2 &    \nodata     &    \nodata     &    \nodata     & 14.70$\pm$.07  & 13.94$\pm$.06  & 12.80$\pm$.07  & 11.49$\pm$.06  &  8.03$\pm$.19 &   \nodata     & GC &  \\
 045050.60$-$704547.3 & 16.54$\pm$.17  & 15.61$\pm$.17  & 15.14$\pm$.18  & 14.48$\pm$.06  & 13.92$\pm$.06  & 14.09$\pm$.11  & 11.59$\pm$.06  &  8.03$\pm$.19 &   \nodata     & GC &  \\
 045148.60$-$671150.7 &    \nodata     &    \nodata     &    \nodata     &    \nodata     & 13.69$\pm$.06  & 12.70$\pm$.08  & 11.59$\pm$.08  &    \nodata    &   \nodata     & GC &  \\
 045201.84$-$664653.4 &    \nodata     &    \nodata     &    \nodata     & 13.58$\pm$.06  & 12.54$\pm$.06  & 11.64$\pm$.06  & 10.50$\pm$.06  &  6.87$\pm$.15 &  2.77$\pm$.24 & GC &  \\

\enddata
\tablecomments{Table~\ref{tab_pgal} is presented in its entirety in the
electronic edition of the Astrophysical Journal.}
\end{deluxetable}

\begin{deluxetable}{lrrrrrrrrrcl}
\rotate
\tablewidth{0pt}
\tabletypesize{\scriptsize}
\tablecaption{Diffuse Non-Sources\label{tab_diff}}
\tablehead{
\colhead{Source ID} & 
\colhead{m$_{J}$} & \colhead{m$_{H}$} & \colhead{m$_{K_{\rm s}}$} &
\colhead{m$_{3.6}$} & \colhead{m$_{4.5}$} & \colhead{m$_{5.8}$} & 
\colhead{m$_{8.0}$} & \colhead{m$_{24.0}$} & \colhead{m$_{70.0}$} &
\colhead{}      & \colhead{Cross} \\
\colhead{ } & 
\colhead{[mag]} & \colhead{[mag]} & \colhead{[mag]} &
\colhead{[mag]} & \colhead{[mag]} & \colhead{[mag]} & 
\colhead{[mag]} & \colhead{[mag]} & \colhead{[mag]} &
\colhead{Class} & \colhead{Identification} 
}
\startdata
 044942.60$-$691232.8 &    \nodata     &    \nodata     &    \nodata     & 14.11$\pm$.09  & 13.63$\pm$.07  & 11.59$\pm$.10  &  9.94$\pm$.09  &    \nodata    &   \nodata     & D &  \\
 044943.58$-$691300.8 &    \nodata     &    \nodata     &    \nodata     & 12.83$\pm$.08  & 12.76$\pm$.08  & 10.39$\pm$.09  &  8.64$\pm$.10  &    \nodata    &   \nodata     & D &  \\
 045205.39$-$665513.8 &    \nodata     &    \nodata     &    \nodata     & 11.88$\pm$.12  & 11.04$\pm$.10  &  9.62$\pm$.13  &  7.74$\pm$.14  &    \nodata    &   \nodata     & D &  \\
 045209.32$-$692015.5 &    \nodata     &    \nodata     &    \nodata     & 14.03$\pm$.08  & 13.95$\pm$.08  & 13.13$\pm$.12  & 11.38$\pm$.13  &    \nodata    &   \nodata     & DC &  \\
 045359.44$-$691001.4 &    \nodata     &    \nodata     &    \nodata     & 13.45$\pm$.09  & 13.69$\pm$.09  & 10.79$\pm$.09  &  9.01$\pm$.12  &    \nodata    &   \nodata     & D &  \\

\enddata
\tablecomments{Table~\ref{tab_diff} is presented in its entirety in the
electronic edition of the Astrophysical Journal.}
\end{deluxetable}

\begin{deluxetable}{lrrrrrrrrrcl}
\rotate
\tablewidth{0pt}
\tabletypesize{\scriptsize}
\tablecaption{Definite LMC Young Stellar Objects\label{tab_yso1}}
\tablehead{
\colhead{Source ID} & 
\colhead{m$_{J}$} & \colhead{m$_{H}$} & \colhead{m$_{K_{\rm s}}$} &
\colhead{m$_{3.6}$} & \colhead{m$_{4.5}$} & \colhead{m$_{5.8}$} & 
\colhead{m$_{8.0}$} & \colhead{m$_{24.0}$} & \colhead{m$_{70.0}$} &
\colhead{}      & \colhead{Cross} \\
\colhead{ } & 
\colhead{[mag]} & \colhead{[mag]} & \colhead{[mag]} &
\colhead{[mag]} & \colhead{[mag]} & \colhead{[mag]} & 
\colhead{[mag]} & \colhead{[mag]} & \colhead{[mag]} &
\colhead{Class} & \colhead{Identification} 
}
\startdata
 044717.50$-$690930.3 & 16.08$\pm$.10  & 15.41$\pm$.13  & 14.44$\pm$.10  & 12.31$\pm$.05  & 11.96$\pm$.05  &  9.64$\pm$.05  &  7.84$\pm$.06  &  2.20$\pm$.11 & -0.82$\pm$.22 & C &  \\
 044733.93$-$703139.2 &    \nodata     &    \nodata     &    \nodata     & 13.62$\pm$.06  & 13.52$\pm$.06  & 11.01$\pm$.06  &  9.31$\pm$.06  &  6.50$\pm$.15 &  0.78$\pm$.22 & C &  \\
 044837.31$-$671834.8 &    \nodata     &    \nodata     &    \nodata     & 13.67$\pm$.06  & 12.68$\pm$.06  & 11.38$\pm$.06  & 10.07$\pm$.06  &  5.54$\pm$.12 &  0.57$\pm$.22 & C &  \\
 044839.94$-$692023.7 &    \nodata     &    \nodata     &    \nodata     & 12.56$\pm$.05  & 11.55$\pm$.05  & 10.24$\pm$.06  &  9.07$\pm$.06  &  5.38$\pm$.12 &  1.30$\pm$.22 & C &  \\
 044847.71$-$691248.3 &    \nodata     &    \nodata     &    \nodata     & 13.85$\pm$.06  & 13.06$\pm$.06  & 12.19$\pm$.07  & 10.94$\pm$.07  &  7.52$\pm$.18 &   \nodata     & C &  \\

\enddata
\tablecomments{Table~\ref{tab_yso1} is presented in its entirety in the
electronic edition of the Astrophysical Journal.}
\tablecomments{Sources marked ``sat.'' at 24~$\mu$m are saturated or amid saturated diffuse emission.}
\end{deluxetable}

\begin{deluxetable}{lrrrrrrrrrcl}
\rotate
\tablewidth{0pt}
\tabletypesize{\scriptsize}
\tablecaption{Probable LMC Young Stellar Objects\label{tab_yso2}}
\tablehead{
\colhead{Source ID} & 
\colhead{m$_{J}$} & \colhead{m$_{H}$} & \colhead{m$_{K_{\rm s}}$} &
\colhead{m$_{3.6}$} & \colhead{m$_{4.5}$} & \colhead{m$_{5.8}$} & 
\colhead{m$_{8.0}$} & \colhead{m$_{24.0}$} & \colhead{m$_{70.0}$} &
\colhead{}      & \colhead{Cross} \\
\colhead{ } & 
\colhead{[mag]} & \colhead{[mag]} & \colhead{[mag]} &
\colhead{[mag]} & \colhead{[mag]} & \colhead{[mag]} & 
\colhead{[mag]} & \colhead{[mag]} & \colhead{[mag]} &
\colhead{Class} & \colhead{Identification} 
}
\startdata
 044609.84$-$664257.7 & 15.91$\pm$.09  & 15.22$\pm$.12  & 14.75$\pm$.13  & 12.73$\pm$.06  & 12.34$\pm$.06  & 10.12$\pm$.06  &  8.31$\pm$.06  &  4.27$\pm$.12 &  0.16$\pm$.22 & CS &   MCBB\,5-24 \\
 044859.94$-$685506.6 & 16.96$\pm$.21  & 15.32$\pm$.12  & 14.20$\pm$.10  & 12.21$\pm$.05  & 11.01$\pm$.05  &  9.91$\pm$.05  &  8.65$\pm$.06  &  4.26$\pm$.11 &  2.14$\pm$.23 & CG &  \\
 044937.07$-$691201.0 &    \nodata     &    \nodata     &    \nodata     & 13.50$\pm$.06  & 13.32$\pm$.06  & 11.17$\pm$.06  &  9.46$\pm$.06  &  6.91$\pm$.16 &   \nodata     & CD &  \\
 045009.44$-$682532.2 &    \nodata     &    \nodata     &    \nodata     & 12.77$\pm$.06  & 12.57$\pm$.07  & 10.15$\pm$.07  &  8.39$\pm$.07  &  5.45$\pm$.12 & -0.53$\pm$.22 & CD &  \\
 045035.36$-$692923.4 & 16.42$\pm$.15  &    \nodata     & 15.25$\pm$.20  & 12.94$\pm$.06  & 12.91$\pm$.06  & 10.18$\pm$.06  &  8.42$\pm$.07  &  5.65$\pm$.13 &   \nodata     & CD &  \\

\enddata
\tablecomments{Table~\ref{tab_yso2} is presented in its entirety in the
electronic edition of the Astrophysical Journal.}
\end{deluxetable}

\begin{deluxetable}{lrrrrrrrrrcl}
\rotate
\tablewidth{0pt}
\tabletypesize{\scriptsize}
\tablecaption{Possible LMC Young Stellar Objects\label{tab_yso3}}
\tablehead{
\colhead{Source ID} & 
\colhead{m$_{J}$} & \colhead{m$_{H}$} & \colhead{m$_{K_{\rm s}}$} &
\colhead{m$_{3.6}$} & \colhead{m$_{4.5}$} & \colhead{m$_{5.8}$} & 
\colhead{m$_{8.0}$} & \colhead{m$_{24.0}$} & \colhead{m$_{70.0}$} &
\colhead{}      & \colhead{Cross} \\
\colhead{ } & 
\colhead{[mag]} & \colhead{[mag]} & \colhead{[mag]} &
\colhead{[mag]} & \colhead{[mag]} & \colhead{[mag]} & 
\colhead{[mag]} & \colhead{[mag]} & \colhead{[mag]} &
\colhead{Class} & \colhead{Identification} 
}
\startdata
 044653.63$-$670109.9 & 16.35$\pm$.15  & 15.61$\pm$.15  & 15.01$\pm$.17  & 13.71$\pm$.06  & 12.85$\pm$.06  & 11.75$\pm$.06  & 10.14$\pm$.05  &  6.09$\pm$.14 &  1.94$\pm$.23 & SC &  \\
 044808.84$-$684219.4 & 16.96$\pm$.20  & 15.64$\pm$.15  &    \nodata     & 13.74$\pm$.06  & 12.78$\pm$.06  & 11.83$\pm$.06  & 10.19$\pm$.06  &  6.47$\pm$.15 &  1.73$\pm$.22 & GC &  \\
 044853.98$-$695856.2 &    \nodata     &    \nodata     &    \nodata     & 14.70$\pm$.07  & 13.94$\pm$.06  & 12.80$\pm$.07  & 11.49$\pm$.06  &  8.03$\pm$.19 &   \nodata     & GC &  \\
 045050.60$-$704547.3 & 16.54$\pm$.17  & 15.61$\pm$.17  & 15.14$\pm$.18  & 14.48$\pm$.06  & 13.92$\pm$.06  & 14.09$\pm$.11  & 11.59$\pm$.06  &  8.03$\pm$.19 &   \nodata     & GC &  \\
 045140.75$-$674456.9 & 14.43$\pm$.04  & 14.16$\pm$.06  & 14.05$\pm$.07  & 13.60$\pm$.06  & 13.30$\pm$.06  & 11.35$\pm$.06  &  8.49$\pm$.06  &  4.35$\pm$.12 &   \nodata     & PSC &  \\

\enddata
\tablecomments{Table~\ref{tab_yso3} is presented in its entirety in the
electronic edition of the Astrophysical Journal.}
\end{deluxetable}

\begin{deluxetable}{lrrrrrrrrrcl}
\rotate
\tablewidth{0pt}
\tabletypesize{\scriptsize}
\tablecaption{Stellar Sources\label{tab_stellar}}
\tablehead{
\colhead{Source ID} & 
\colhead{m$_{J}$} & \colhead{m$_{H}$} & \colhead{m$_{K_{\rm s}}$} &
\colhead{m$_{3.6}$} & \colhead{m$_{4.5}$} & \colhead{m$_{5.8}$} & 
\colhead{m$_{8.0}$} & \colhead{m$_{24.0}$} & \colhead{m$_{70.0}$} &
\colhead{}      & \colhead{Cross} \\
\colhead{ } & 
\colhead{[mag]} & \colhead{[mag]} & \colhead{[mag]} &
\colhead{[mag]} & \colhead{[mag]} & \colhead{[mag]} & 
\colhead{[mag]} & \colhead{[mag]} & \colhead{[mag]} &
\colhead{Class} & \colhead{Identification} 
}
\startdata
 042728.61$-$694516.0 &  6.88$\pm$.02  &  6.38$\pm$.03  &  6.25$\pm$.02  &  6.13$\pm$.05  &  8.41$\pm$.06  &  6.21$\pm$.05  &  6.12$\pm$.05  &    \nodata    &  $>$3.88      & S &   HD\,28898 \\
 043849.02$-$692715.8 &    \nodata     &    \nodata     &    \nodata     &    \nodata     &  8.64$\pm$.09  &    \nodata     &  6.29$\pm$.06  &    \nodata    &   \nodata     & S &   HD\,30083 \\
 043956.28$-$703913.9 & 15.90$\pm$.07  & 15.18$\pm$.08  & 14.95$\pm$.12  & 14.41$\pm$.06  & 13.83$\pm$.06  & 13.05$\pm$.07  & 11.66$\pm$.06  &  9.11$\pm$.26 &   \nodata     & S &  \\
 044302.05$-$703405.5 & 14.90$\pm$.07  & 14.45$\pm$.09  & 13.95$\pm$.10  & 11.42$\pm$.06  & 11.19$\pm$.06  &  8.95$\pm$.08  &  7.05$\pm$.06  &  2.48$\pm$.11 & -2.14$\pm$.22 & S &   MSX\,LMC\,1134 \\
 044354.00$-$695607.7 &    \nodata     &    \nodata     &    \nodata     & 12.84$\pm$.05  & 11.79$\pm$.05  & 10.84$\pm$.06  &  9.71$\pm$.05  &  5.98$\pm$.13 &   \nodata     & SG &  \\

\enddata
\tablecomments{Table~\ref{tab_stellar} is presented in its entirety in the
electronic edition of the Astrophysical Journal.}
\tablecomments{Sources marked ``sat.'' at 24~$\mu$m are saturated or amid saturated diffuse emission.}
\end{deluxetable}

\begin{deluxetable}{lccc}
\tablewidth{0pt}
\tabletypesize{\footnotesize}
\tablecaption{Summary of Sources Classification from Our Catalog\label{tab_summary_class}}
\tablehead{
\colhead{Class} & \colhead{Total} & \colhead{m$_{8.0\mu{\rm m}} <$8.0} & \colhead{Possible YSOs} 
}
\startdata
Definite YSOs &  855 & 233 & \nodata \\
Probable YSOs &  317 &  14 & \nodata \\
Possible YSOs &  213 &     &   213   \\
AGB/post-AGB  &  117 & 111 &     0   \\
PNe           &   56 &   9 &     2   \\
Galaxy        & 1075 &   7 &   116   \\
Diffuse       &  159 &  12 &    44   \\
Stellar       &  291 &  45 &    51   \\ 
Reject        &   40 &  11 & \nodata 
\enddata
\end{deluxetable}

\begin{deluxetable}{lccccccccc}
\tablewidth{0pt}
\tabletypesize{\footnotesize}
\tablecaption{Comparison between SAGE and Our Catalog\label{tab_sagecomp1}}
\tablehead{
\multicolumn{2}{c}{SAGE} & \colhead{Sources} & \colhead{Sources} & \colhead{Sources} & \multicolumn{5}{c}{} \\
\multicolumn{2}{c}{Results} & \colhead{Among} & \colhead{Not In} & \colhead{Not in} & \multicolumn{5}{c}{Reason Not In Our Sample} \\
\colhead{}      & \colhead{}      & \colhead{Our}   & \colhead{Our} & \colhead{Our} & 
\colhead{AGB}   & \colhead{Galaxy} & \colhead{both} & \colhead{no} & \colhead{no} \\
\colhead{Class} & \colhead{Total} & \colhead{Sample\tablenotemark{a}} & \colhead{Sample\tablenotemark{a}} & \colhead{Catalog\tablenotemark{b}} &
\colhead{cutoff} & \colhead{cutoff} & \colhead{cutoffs} & \colhead{4.5$\mu$m} & \colhead{8.0$\mu$m} 
}
\startdata
YSO\_hp & 458 & 326 & 131 & 1 & 122 &   2 &  1 &  8 &  0 \\
YSO     & 532 & 161 & 366 & 5 & 139 & 184 &  5 & 32 & 16 \\
Evolved & 118 &  41 &  76 & 1 &  69 &   5 &  2 &  0 &  4 \\
PN      &  82 &  49 &  33 & 0 &  12 &  20 &  3 &  2 &  2 \\
Galaxy  &   4 &   1 &   3 & 0 &   3 &   0 &  0 &  0 &  0 \\
other   &   3 &   1 &   2 & 0 &   1 &   0 &  0 &  1 &  0 \\
\enddata
\tablenotetext{a}{Sources in our``sample" refer to all sources from the 
wedge in the [8.0] vs. [4.5]$-$[8.0] CMD.}
\tablenotetext{b}{Sources in our``catalog" refer to all sources identified
in our initial photometric measurements.}
\end{deluxetable}

\begin{deluxetable}{lccccccccc}
\tablewidth{0pt}
\tabletypesize{\footnotesize}
\tablecaption{Comparison between SAGE and Our Classification\label{tab_sagecomp2}}
\tablehead{
\colhead{}     & \colhead{Sources} & \multicolumn{7}{c}{} \\
\colhead{}     & \colhead{Among} & \multicolumn{7}{c}{Our Class} \\
\colhead{SAGE} & \colhead{Our} & \colhead{Definite} & \colhead{Probable} & 
  \colhead{Possible} & \colhead{} & \colhead{Background} &
  \colhead{Evolved} & \colhead{} & \colhead{} \\
\colhead{Class} & \colhead{Sample\tablenotemark{a}} & \colhead{YSO} & \colhead{YSO} & \colhead{YSO} & 
\colhead{GC} & \colhead{Galaxy} & \colhead{Star} & \colhead{ERO} & \colhead{PN} 
}
\startdata
YSO\_hp & 326 & 156 & 51 & 41 & 31 & 65 &  6 & 7 &  0 \\
YSO     & 161 &  86 & 13 & 13 &  9 & 29 & 19 & 1 &  0 \\
Evolved &  41 &   1 &  3 &  2 &  0 &  0 & 35 & 0 &  0 \\
PN      &  49 &   6 &  5 &  1 &  1 &  0 &  3 & 0 & 34 \\
Galaxy  &   1 &   0 &  0 &  1 &  1 &  0 &  0 & 0 &  0 \\
\enddata
\tablenotetext{a}{Sources in our``sample" refer to all sources from the 
wedge in the [8.0] vs. [4.5]$-$[8.0] CMD.}
\end{deluxetable}

\begin{deluxetable}{ccccc}
\tablewidth{0pt}
\tablecaption{YSO Luminosity Function Fit Results\label{tab_lumfunc}}
\tablehead{
\colhead{Band} & \colhead{$a$} & \colhead{$\sigma_a$} & \colhead{$a$\tablenotemark{1}} & \colhead{$\sigma_a$\tablenotemark{1}} 
}
\startdata
3.6 & 0.43 & 0.02 & 0.39 & 0.03 \\
4.5 & 0.35 & 0.02 & 0.35 & 0.03 \\
5.8 & 0.40 & 0.04 & 0.41 & 0.04 \\
8.0 & 0.44 & 0.02 & 0.45 & 0.02 \\
\enddata
\tablenotetext{1}{Values for $a$ and $\sigma_a$ for a hypothetical luminosity function 
using our YSOs and those from the SAGE analysis with $[4.5]-[8.0]<2.0$}
\end{deluxetable}

\end{document}